\documentclass[sigconf]{acmart}
\usepackage{graphicx}

\usepackage{ifthen}
\newboolean{techreport}
\usepackage{caption}
\usepackage{multirow}
\usepackage{subcaption}
\usepackage{hyperref}
\usepackage{amsmath}
\usepackage{tikz}
\usepackage[]{footmisc}
\usepackage{graphicx}% http://ctan.org/pkg/graphicx
\usepackage{pifont}% http://ctan.org/pkg/pifont

\usepackage[ruled,linesnumbered,vlined]{algorithm2e}
\usepackage{amsmath,amstext}
\let\oldnl\nl% Store \nl in \oldnl
\newcommand{\nonl}{\renewcommand{\nl}{\let\nl\oldnl}}% Remove line number for one line
\usepackage{tikz}

\definecolor{codegreen}{rgb}{0,0.6,0}
\definecolor{codegray}{rgb}{0.5,0.5,0.5}
\definecolor{codepurple}{HTML}{C42043}
\definecolor{backcolour}{rgb}{1,1,1}

\newcommand\numberstyle[1]{%
    \footnotesize
    \color{codegray}%
    \ttfamily
    \ifnum#1<10 0\fi#1 |%
}

\newcommand*\circled[1]{\tikz[baseline=(char.base)]{
    \node[shape=circle,draw,inner sep=0.2pt] (char) {#1};}}

\newtheorem{definition}{Definition}[section]

\usepackage{enumitem}
\renewcommand\footnotetextcopyrightpermission[1]{} % removes footnote with conference information in first column

\begin{document}
\setboolean{techreport}{true}
% ****************** TITLE ****************************************

\title{Nova-LSM: A Distributed, Component-based LSM-tree Key-value Store}
\subtitle{Extended Version}
\titlenote{A shorter version of this paper appeared in ACM-SIGMOD 2021~\cite{novalsm-sigmod}.}
\author{Haoyu Huang}
\affiliation{%
  \institution{University of Southern California}
    \city{Los Angeles}
  \country{USA}
}
\email{haoyuhua@usc.edu}

\author{Shahram Ghandeharizadeh}
\affiliation{%
  \institution{University of Southern California}
    \city{Los Angeles}
  \country{USA}
}
\email{shahram@usc.edu}

\begin{abstract}
The cloud infrastructure motivates disaggregation of monolithic data stores into components that are assembled together based on an application's workload.  This study investigates disaggregation of an LSM-tree key-value store into components that communicate using RDMA.  These components separate storage from processing, enabling processing components to share storage bandwidth and space.  The processing components scatter blocks of a file (SSTable) across an arbitrary number of storage components and balance load across them using power-of-d.  They construct ranges dynamically at runtime to parallelize compaction and enhance performance.  Each component has configuration knobs that control its scalability.  The resulting component-based system, Nova-LSM, is elastic.  It outperforms its monolithic counterparts, both LevelDB and RocksDB, by several orders of magnitude with workloads that exhibit a skewed pattern of access to data.

\end{abstract}

\maketitle
\pagestyle{plain} % removes running headers

\section{Introduction}\label{sec:intro}
Persistent multi-node key-value stores (KVSs) are used widely by diverse applications.  This is due to their simplicity, high performance, and scalability.  Example applications include online analytics~\cite{hbase-airbnb,hbase-flurry}, product search and recommendation~\cite{hbase-alibaba}, graph storage~\cite{dragon,zen,myrocks}, hybrid transactional and analytical processing systems~\cite{f1-lightening}, and finance~\cite{rocksdb}.  For write-intensive workloads, many KVSs implement principles of the Log-Structured Merge-tree (LSM-tree)~\cite{leveldb,rocksdb,bigtable,cassandra,hbase}.  
% \textbf{Prof:Removed.}  
% The key insight of LSM-tree is that a sequential disk I/O is significantly faster than a random I/O by avoiding the overhead of seek and rotational latency.
%%%Hence, it transforms a random I/O into multiple sequential I/Os by not performing in-place updates.  
%%%Moreover, it uses in-memory memtables to buffer writes and subsequently flushes them as Sorted String Tables (SSTables) to disk using sequential I/Os.  
They do not perform updates in place and use in-memory memtables to buffer writes and flush them as Sorted String Tables (SSTables) using sequential disk I/Os.

To improve resource utilization, cloud-native data stores leverage the cloud infrastructure with hardware-software co-design to enable elasticity~\cite{aurora,socrates,Taurus}. 
This motivates disaggregation of monolithic data stores into components~\cite{nova,gamma90,novalsm-thesis}.
Logically, components may separate storage from processing.  
Physically, a component utilizes the appropriate hardware and scales independent of the others.  
Fast networking hardware with high bandwidths, e.g., Remote Direct Memory Access (RDMA)~\cite{azure-hpc}, provides connectivity for data exchange between components that constitute a data store.

We present Nova-LSM, a distributed, component-based LSM-tree data store. 
Its components include LSM-tree Component (LTC), Logging Component (LogC), and Storage Component (StoC).  
Both LTC and LogC use StoC to store data.  
LogC uses StoC to enhance either availability or durability of writes.
Nova-LSM enables each component to scale independent of the others. 
With an I/O intensive workload, a deployment may include a few LTCs and LogCs running on compute-optimized servers with hundreds of StoCs running on storage-optimized servers; see~\cite{ec2-instance-types,azure-hpc, legoos} for such servers.

An application range partitions data across LTCs.
Physically, an LTC may organize memtables and SSTables of a range in different ways:
assign a range and its SSTables to one StoC, 
assign SSTables of a range to different StoCs with one SSTable assigned to one StoC, or assign SSTables of a range to different StoCs and scatter blocks of each SSTable across multiple StoCs.  
An LTC may use either replication, a parity-based technique or a hybrid of the two to enhance availability of data in the presence of StoC failures.
%%%~\cite{gfs,socrates}. 
%%%We focus on one replica in this study. 

% The configuration setting of an LTC dictates which of the above 3 choices are employed by its deployment.

A LogC may be a standalone component or a library integrated with an LTC.  
It may be configured to support availability, durability, or both.
%%%It may use StoC to implement alternative choices. 
Availability is implemented by replicating log records in memory, providing the fastest service times.  
Durability is implemented by persisting log records at StoCs.
%With both, LogC enhances availability by maintaining recent log records in memory to enhance mean time to repair.  

%%%It is possible for a LogC to use the memory of LTCs or standalone components that provide memory to implement its functionality.  
%%%In this study, we assume LogC is integrated with LTC.  

A StoC is a simple component that stores, retrieves, and manages variable-sized blocks.  
%%It may support either a single-level or a hierarchy of storage devices.  
Its speed depends on the choice of storage devices and whether they are organized in a hierarchy.
An example hierarchy may consist of SSD and disk, using a write-back policy to write data from SSD to disk asynchronously~\cite{flashcache,bluecache,fb-flashcache}.

\begin{figure}[!ht]
    \centering
    \includegraphics[width=\linewidth]{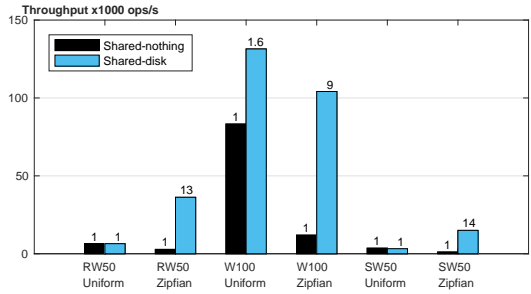}
    \caption{Performance comparison between shared-nothing and shared-disk with a 1 TB database range partitioned across 10 servers.}
    \label{fig:intro-performance}
\end{figure}

Nova-LSM uses RDMA to enable LTCs to share disk bandwidth of StoCs.  Figure~\ref{fig:intro-performance} shows the performance benefits of this architecture when compared with the classical shared-nothing architecture~\cite{sharedNothing,gamma90}.
It presents throughput of 3 different YCSB~\cite{ycsb} workloads with (1) a uniform access pattern to data and (2) after a sudden change to a skewed pattern of access to data.  These are labeled Uniform and Zipfian on the x-axis, respectively.  
%%%Our target database is 1 TB and partitioned across 10 nodes.  The disk bandwidth is the bottleneck with all workloads and access patterns.  
We show throughput of two configurations of Nova-LSM using the same hardware platform consisting of 10 nodes.  The one labeled shared-disk uses RDMA to enable LTCs to share disk bandwidth and space of all ten StoCs.  With the shared-nothing, an LTC stores its SSTables on the StoC local to its node.  The number on each bar is the factor of improvement in throughput with the shared-disk when compared with the shared-nothing.  

With Zipfian, the shared-disk improves performance by a factor of 9 or higher.  The shared-nothing directs most of the requests to the StoC with the most popular data item, exhausting its disk bandwidth while nine StoCs wait for work with a disk bandwidth utilization lower than 20\%.  By requiring LTCs to share StoCs, Nova-LSM distributes the imposed load more evenly across all 10 disks, improving system throughput dramatically with Zipfian.

\subsection{Challenges and Solutions}\label{sec:intro:challenges}
\noindent {\bf Challenge 1:}  A challenge of LSM-tree implementations including LevelDB~\cite{leveldb}, RocksDB~\cite{rocksdb,myrocks}, and AsterixDB~\cite{asterixdb} is {\em write stalls}~\cite{lsm-write-stalls,bLSM,silk,matrixkv2020}.  It refers to moments in time when the system suspends processing of writes because either all memtables are full or the size of SSTables at Level$_0$ exceeds a threshold.  
In the first case, writes are resumed once an immutable memtable is flushed to disk.  
In the second, writes are resumed once compaction of SSTables at Level$_0$ into Level$_1$ causes the size of Level$_0$ SSTables to be lower than its pre-specified threshold.  

\noindent{\bf Solution 1:}  Nova-LSM addresses this challenge using both its architecture that separates storage from processing and a divide-and-conquer approach.  Consider each in turn.  

\begin{figure}[!t]
    \centering
    \includegraphics[width=\linewidth]{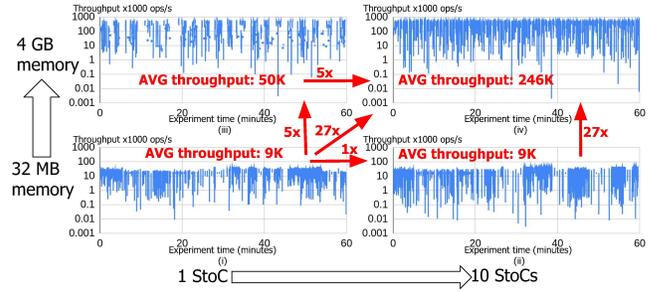}
    \caption{Throughput of different Nova-LSM configurations.}
    \label{fig:write-stall-intro}
\end{figure}

With Nova-LSM's architecture, one may either increase the amount of memory allocated to an LTC, the number of StoCs to increase the disk bandwidth, or both. 
Figure~\ref{fig:write-stall-intro} shows the throughput observed by an application for a one-hour experiment with different Nova-LSM configurations.
The base system consists of one LTC with 32 MB of memory and 1 StoC, see Figure~\ref{fig:write-stall-intro}.i.
As a function of time, its throughput spikes and drops down to zero due to write stalls, providing an average throughput of 9,000 writes per second.  
The log-scale for the y-axis highlights the significant throughput variation since writes must wait for memtables to be flushed to disk.   
Increasing the number of StoCs to 10 does not provide a benefit (see Figure~\ref{fig:write-stall-intro}.ii) because the number of memtables is too few. 
However, increasing the number of memtables to 128 enhances the peak throughput from ten thousand writes per second to several hundred thousand writes per second, see Figure~\ref{fig:write-stall-intro}.iii.
It improves the average throughput 5 folds to 50,000 writes per second and utilizes the disk bandwidth of 1 StoC fully.
Write stalls continue to cause the throughput to drop to close to zero for a noticeable amount of time, resulting in a sparse chart. 
Finally, increasing the number of StoCs of this configuration to 10 diminishes write stalls significantly, see Figure~\ref{fig:write-stall-intro}.iv.
The throughput of this final configuration is 27 folds higher than the throughput of the base configuration of Figure~\ref{fig:write-stall-intro}.i.

Nova-LSM reduces complexity of software to implement compaction using a divide and conquer approach by constructing dynamic ranges, {\em Dranges}.
These Dranges are independent of the application specified ranges.  
An LTC creates them with the objective to balance the load imposed by application writes across them.
This enables parallel compactions to proceed independent of one another, utilizing resources that would otherwise sit idle. (Section~\ref{sec:dranges})

\noindent{\bf Challenge 2:}  Increasing the number of memtables (i.e., amount of memory) slows down scans and gets by requiring them to search a larger number of memtables and SSTables at Level$_0$.

\noindent{\bf Solution 2:}
Nova-LSM implements an index that identifies either the memtable or the SSTable at Level$_0$ that stores the latest value of a key. 
If a get observes a hit in the lookup index, it searches either one memtable or one SSTable at Level$_0$. 
Otherwise, it searches SSTables at higher levels that overlap with the referenced key. 
This index improves the throughput by a factor of 1.7 (2.8) with a uniform (skewed) pattern of access to data.  (Section~\ref{sec:ltc:gets})

Dranges (see Solution 1) partition the keyspace to enable an LTC to process a scan by searching a subset of memtables and SSTables. 
This improves the throughput 26 (18) folds with a uniform (skewed) pattern of access to data. (Section~\ref{sec:ltc:scans})

\noindent{\bf Challenge 3:}  When LTCs assign SSTables to StoCs randomly, temporary bottlenecks may form due to random collision of many writes to a few, potentially one, StoC.
This slows down requests due to queuing delays. 

\noindent{\bf Solution 3:}
Nova-LSM addresses this challenge in two ways.
First, it scatters blocks of a SSTable across multiple StoCs.
Second, it uses power-of-d~\cite{power-of-two} to assign writes to StoCs.
Consider each in turn. 
An LTC may partition a SSTable into $\rho$ fragments and use RDMA to write these fragments to $\rho$ StoCs.
When the value of $\rho$ equals the total number of StoCs, $\beta$, the possibility of random collisions is eliminated altogether.
For example with $\beta$=10 StoCs, the throughput doubles when an LTC scatters a SSTable across $\rho$=10 StoCs instead of one StoC. 

Full partitioning is undesirable with large values of $\beta$.
It wastes disk bandwidth by increasing the overall number of disk seeks and rotational latencies to write a SSTable.  Hence, Nova-LSM supports partitioning degrees lower than the number of StoCs, $\rho < \beta$.  (Section~\ref{sec:scatter})

With $\rho < \beta$, each LTC implements power-of-d to write a SSTable as follows.  
It peeks at the disk queue sizes of $\rho*2$ randomly selected StoCs, writing to those $\rho$ StoCs with the shortest disk queues.  Experimental results of Figure~\ref{fig:intro-performance} use $\rho$=3 with $\beta$=10.  Note that power-of-d enables the write-only workload W100 with a uniform access pattern to observe a 60\% improvement.  It minimizes the impact of transient bottlenecks attributed to random independent requests colliding on one StoC.  
\newpage
\subsection{Contributions and Outline}\label{sec:intro:outline}
The primary contributions of this study are as follows. 
\begin{itemize}[leftmargin=*]
    \item An LTC constructs dynamic ranges to enable parallel compactions for SSTables at Level$_0$, utilizing resources that would otherwise sit idle and minimizing the duration of write stalls.   (Section~\ref{sec:ltc})
    \item An LTC maintains indexes to provide fast lookups and scans. (Section~\ref{sec:ltc})
    \item An LTC scatters blocks of one SSTable across a subset of StoCs (e.g., 3 out of $\beta > 3$) while scattering blocks of different SSTables across all $\beta$ StoCs. It uses the power-of-d random choices~\cite{power-of-two} to minimize queuing delays at StoCs. An LTC implements replication and a parity-based technique to enhance data availability in the presence of StoC failures. (Section~\ref{sec:component-interfaces} and~\ref{sec:ltc})
    \item A LogC separates the availability of log records from their durability. LogC configured with availability using RDMA recovers 4 GB of log records within one second. (Section~\ref{sec:logc})
    \item A StoC implements simple interfaces that allow it to scale horizontally. (Section~\ref{sec:stoc})
    \item A highly elastic Nova-LSM.  (Section~\ref{sec:elasticity})
    \item A publicly available implementation of  Nova-LSM as an extension of LevelDB\footnote{\url{https://github.com/HaoyuHuang/NovaLSM}}. Nova-LSM shows a 2.3x to 18x higher throughput than LevelDB and RocksDB with a skewed access pattern and a 2 TB database.  Nova-LSM scales vertically as a function of memory sizes. It also scales horizontally as a function of LTCs and StoCs. (Section~\ref{sec:eval})
    \item The main limitation of Nova-LSM is its higher CPU utilization.  With 1 LTC and 1 StoC running on 1 server, this is attributed to the overhead of maintaining the index that addresses Challenge 2.  With additional LTCs, this overhead extends to include pulling of messages by the RDMA threads.  Total overhead results in a 20-30\% performance degradation with a CPU intensive workload (50\% scan, 50\% write) using a uniform pattern of access to data.  With the same workload and a skewed access pattern, Nova-LSM outperforms LevelDB by a factor of 5. (Figure~\ref{fig:eval:nova-leveldb-100gb} of Section~\ref{sec:eval:leveldb-large-db})
\end{itemize}

The rest of this paper is organized as follows. 
Section~\ref{sec:bg} presents the background on LevelDB and RDMA. 
We then present the architecture of Nova-LSM and its component interfaces in Section~\ref{sec:nova-lsm}. 
Sections~\ref{sec:ltc} to~\ref{sec:stoc} detail the design of LTC, LogC, and StoC, respectively. 
Section~\ref{sec:impl} details the implementation of Nova-LSM.
We present our evaluation results in Section~\ref{sec:eval}. 
Section~\ref{sec:elasticity} presents the elasticity of Nova-LSM. 
Section~\ref{sec:related-work} surveys related work. Section~\ref{sec:conclusion} concludes and discusses future research directions.

\section{Background}\label{sec:bg}
\subsection{LevelDB}\label{sec:bg:leveldb}
LevelDB organizes key-value pairs in two memtables and many immutable Sorted String Tables (SSTables).
While the memtables are main memory resident, SSTables reside on disk.  
Keys are sorted inside a memtable and a SSTable based on the application specified comparison operator. 
% Memtables are append-only and SSTables are immutable. 
LevelDB organizes SSTables on multiple levels. 
Level$_0$ is semi-sorted while the other levels are sorted.
The expected size of each level is limited and is usually increasing by a factor of 10. 
This bounds the storage size. 
When a memtable becomes full, it is converted to a SSTable at Level$_0$ and written to disk. 
Other data stores built using LevelDB support a configurable number of memtables, e.g., RocksDB~\cite{rocksdb}.

With large memory, it is beneficial to have many small memtables instead of a few large ones.  
LevelDB implements a memtable using skiplists, providing O(log n) for both lookups and inserts where n is the number of keys in the memtable.  
A larger memtable size increases the number of keys and slows down lookups and inserts~\cite{flodb}.  
Typically, an application configures the memtable size to be a few MB, e.g., 16 MB.

\noindent\textbf{Write:} 
A write request is either a put or a delete. 
LevelDB maintains a monotonically increasing sequence number to denote the version of a key.  
A write increments the sequence number and appends the key-value pair associated with the latest value of the sequence number to the active memtable. 
Thus, the value of a key is more up-to-date if it has a greater sequence number.
When a write is a delete, the value of the key is a tombstone. 
When the active memtable is full, LevelDB marks it as immutable and creates a new active memtable to process writes. 
A background compaction thread converts the immutable memtable to a SSTable at Level$_0$ and flushes it to disk. 

\noindent\textbf{Logging:} LevelDB associates a memtable with a log file to implement durability of writes. 
A write appends a log record to the log file prior to writing to the active memtable.
When a compaction flushes the memtable to disk, it deletes the log file. 
In the presence of a failure, during recovery, LevelDB uses the log files to recover its corresponding memtables. 

\noindent\textbf{Get:} A get for a key searches memtables first, then SSTables at Level$_0$, followed by SSTables at higher levels for a value. 
It terminates once the value is found. 
It may search \textbf{all} SSTables at Level$_0$ if each spans the entire keyspace. 
This is undesirable especially when we have a large number of memtables. 
For SSTables at Level$_1$ and higher, it often searches only one SSTable since all keys are sorted. 
Bloom filters may be used to detect when this search may not be necessary.

\noindent\textbf{Compaction:} 
LevelDB uses leveled compaction~\cite{lsmtree-survey} to reduce its disk space usage to remove older versions of key-value pairs in the background. 
It compacts SSTables at Level$_i$ with the highest ratio of actual size to expected size. 
It picks a subset of SSTables at Level$_i$ and computes their overlapping SSTables at Level$_{i+1}$.
Then, it reads these SSTables into memory and compacts them into a new set of SSTables to Level$_{i+1}$. 
This ensures keys are still sorted at Level$_{i+1}$ after compaction. 
LevelDB deletes the compacted SSTables to free disk space. 
With a large number of SSTables at Level$_0$, the compaction may require a long time to complete. 
This is the write stall described in Section~\ref{sec:intro:challenges}.

\subsection{RDMA}
Similar to the Ethernet-based IP Networks, RDMA consists of a switch and a network interface card (RNIC) that plugs into a server.
It provides bandwidths in the order of tens and hundreds of Gbps with latencies in the order of a few microseconds~\cite{rdma-eval}.
The ones used in this study provide a bandwidth of 56 Gbps, see~\cite{cloudlab} for their detailed specifications.
RDMA is novel because it provides one-sided RDMA READ/WRITE verbs that read (write) data from (to) a remote server while bypassing its CPU. 

Two servers are connected using a pair of queues (QP):
QP$_i$ on server $i$ and QP$_j$ on server $j$. 
Its communication is asynchronous. 
A QP may be reliable connected, unreliable connected, or unreliable datagram. 
We focus on reliable connected QPs since the other types may drop packets silently. 
A QP supports three communication verbs: RDMA SEND, RDMA READ, and RDMA WRITE. 
RDMA READ and WRITE bypass the receiver's CPU. 
An RDMA READ reads a remote memory region into its local memory region. 
An RDMA WRITE writes a local memory region to a remote memory region directly.
QP$_i$ may issue a request to QP$_j$ using RDMA SEND.
QP$_j$ notifies server $j$ once it receives the request.
It also generates an acknowledgment to QP$_i$. 
QP$_i$ may tag a request with a 4-byte immediate data. 
In this case, QP$_j$ notifies server $j$ of the immediate data when it receives an RDMA SEND or RDMA WRITE request.

\begin{table}
\small
\centering
\caption{Notations and their definitions.}
\begin{tabular}{ll} \hline
Notation & Definition \\ \hline
$\eta$ & Total number of LTCs. \\ \hline
$\beta$ & Total number of StoCs. \\ \hline
$\omega$ & Number of ranges per LTC. \\ \hline
$\theta$ & Number of Dranges per range. \\ \hline
$\gamma$ & Number of Tranges per Drange. \\ \hline
$\alpha$ & Number of active memtables per range. \\ \hline
$\delta$ & Number of memtables per range. \\ \hline
$\tau$ & Size of memtable/SSTable in MB. \\ \hline
$\rho$ & Number of StoCs to scatter a SSTable. \\  \hline
\end{tabular}
\label{tab:term}
\end{table}
\section{Nova-LSM Components}\label{sec:nova-lsm}
\begin{figure}[!ht]
    \centering
    \includegraphics[width=0.9\linewidth]{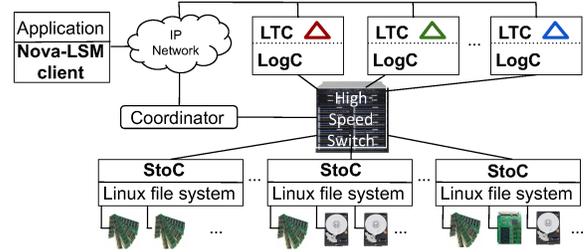}
    \caption{Architecture.}
    \label{fig:arch}
\end{figure}

Figure~\ref{fig:arch} shows the architecture of Nova-LSM, consisting of LSM-tree components (LTCs), logging components (LogCs), and storage components (StoCs). 
This architecture separates storage from processing similar to systems such as Aurora~\cite{aurora}, Socrates~\cite{socrates}, SolarDB~\cite{solardb}, Tell~\cite{tell}, and RockSet~\cite{rockset}.
Table~\ref{tab:term} provides the definition of notations used in this paper.

An LTC consists of $\omega$ ranges.  
The LTC constructs a LSM-tree for each range. 
It processes an application's requests using these trees. 
A LogC is a library integrated into an LTC and is responsible for generating log records to StoCs during normal mode of operation and fetching log records during recovery mode.
A StoC stores, retrieves, and manages variable-sized blocks using alternative storage mediums, e.g., main memory, disk, SSD, or a storage hierarchy.
These components are interconnected using RDMA.

One may partition a database into an arbitrary number of ranges and assign these ranges to $\eta$ LTCs.  In its simplest form, one may partition a database
across $\omega*\eta$ ranges and assign $\omega$ ranges to each LTC. 
The coordinator maintains a configuration that contains the partitioning information and the assignment of ranges to $\eta$ LTCs.
Nova-LSM clients use this configuration information~\cite{rejig,gemini} to direct a request to an LTC with relevant data.
\begin{figure}[!ht]
    \centering
    \includegraphics[width=1.0\linewidth]{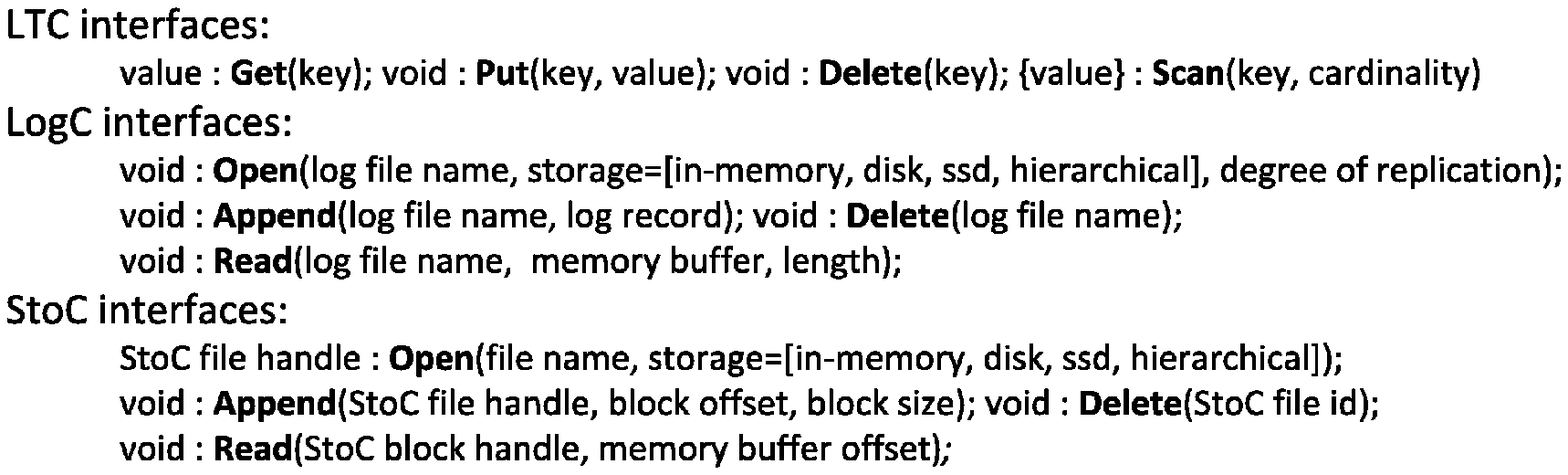}
    \caption{Component Interfaces.}
    \label{fig:interfaces}
\end{figure}

Figure~\ref{fig:arch} shows an IP network for Nova-LSM clients to communicate with LTCs and the coordinator.  This network may be RDMA when the application nodes are in the same data center as the LTCs and the coordinator~\cite{l5}.

Nova-LSM uses mechanisms similar to the Google File System~\cite{gfs} to handle failures.  The coordinator of Figure~\ref{fig:arch} maintains a list of StoCs, LTCs, and ranges assigned to LTCs.  We use leases to minimize management overhead at the coordinator.  A lease has an adjustable initial timeout.  The coordinator grants a lease on a range to an LTC to process requests referencing key-value pairs contained in that range.  Similarly, the coordinator grants a lease to a StoC to process requests.  Both StoC and LTC may request and receive lease extensions from the coordinator indefinitely.  These extension requests and grants are piggybacked on the heartbeat messages regularly exchanged between the coordinator and StoCs/LTCs.  If the coordinator loses communication with an LTC, it may safely grant a new lease on the LTC's assigned ranges to another LTC after the old lease expires.  A StoC/LTC that fails to renew its lease by communicating with the coordinator stops processing requests.
One may use Zookeeper to realize a highly available coordinator~\cite{zookeeper}.

%%%%\textbf{Prof: I replaced the Zookeeper paragraph with 'One may use Zookeeper to realize a highly available coordinator~\cite{zookeeper}'.}
% Zookeeper~\cite{zookeeper} enables a leader coordinator and its followers to replicate the metadata about configurations and leases.   Its messaging layer takes care of replacing the leader on failures and synchronizing followers with leaders.  Moreover, the atomic aspect of the messaging layer guarantees that the local replicas never diverge.

While SSTables are immutable, compactions mutate manifest files containing metadata about levels and their assigned SSTables.  Replicas of a manifest file may become stale if a StoC fails and misses these mutations while it is unavailable.  For each manifest file, Nova-LSM maintains a version number to distinguish between up-to-date and stale manifest replicas~\cite{rejig,gemini}.  If a StoC is unavailable then the version numbers of its manifest replicas are not advanced.  When the StoC restarts, it reports its set of manifest replicas and their version numbers to the coordinator.  The coordinator detects the stale manifest replicas and deletes them.

% There are several advantages of this architecture over today's LevelDB. 
% First, it enables LTCs and LogCs to balance load across different StoCs in the presence of data skew and write bursts.
% Second, each component may be scaled independently of the others.
% Hence, scaling LTCs (i.e., compute resources) does not require data migration at StoC.
% Third, it facilitates elasticity as it decouples the compute from storage. 
% Finally, one may configure each component differently, e.g., one LTC may use replication while another may use a parity based technique for data availability\footnote{Controlled using a configuration file by the coordinator.}.

\begin{figure}[!ht]
    \centering
    \includegraphics[width=0.8\linewidth]{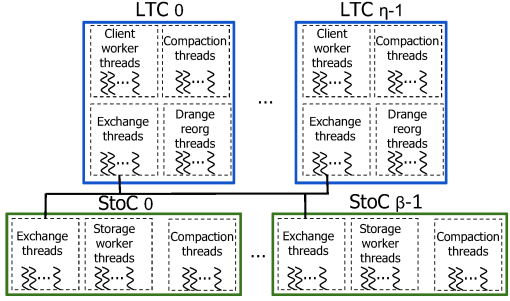}
    \caption{Thread Model.}
    \label{fig:thread}
\end{figure}

\subsection{Component Interfaces}\label{sec:component-interfaces}
LTC provides the standard interfaces of an LSM-tree as shown in Figure~\ref{fig:interfaces}. 
For high availability, one may configure LTC to either replicate a SSTable across StoCs, use a parity-based technique~\cite{raid}, or a combination of the two. 
With parity, a SSTable contains a parity block computed using the $\rho$ data block fragments. 
Once a StoC fails and an LTC references a data fragment on this StoC, the LTC reads the parity block and the other $\rho-1$ data block fragments to recover the missing fragment.
It may write this fragment to a different StoC for use by future references.

LogC provides interfaces for append-only log files. 
It implements these interfaces using StoCs.
For high availability, a LogC client may open a log file with in-memory storage and 3 replicas.

StoC provides variable-sized block interfaces for append-only files. 
A StoC file is identified by a globally unique file id. 
When a StoC client opens a file, StoC allocates a buffer in its memory and returns a file handle to the client.
The file handle contains the file id, memory buffer offset, and its size. 
A StoC client appends a block by writing to the memory buffer of the StoC file directly using RDMA WRITE. 
% A StoC block handle contains the StoC file id, block offset, and block size. 
When a StoC client, say an LTC, reads a block, it first allocates a buffer in its local memory.  Next, it instructs the StoC to write the block to its buffer using RDMA WRITE. 
Sections~\ref{sec:ltc} to~\ref{sec:stoc} detail the design of these interfaces. 

\subsection{Thread Model}\label{sec:thread-model}
Recent studies have shown that RDMA scales poorly with a high number of QPs~\cite{rdma-eval,rdma-guideline}.
Thus, we configure each node with a number of dedicated exchange (xchg) threads to minimize the number of QPs. 
An xchg thread $k$ at node $i$ maintains a QP that connects to the xchg thread $k$ at each of the other nodes. 
Its sole responsibility is to issue requests from its caller to another node and transmit the received responses back to its caller.
It delegates all actual work to other threads. 

LTC has four types of threads, see Figure~\ref{fig:thread}. 
A client worker thread processes client requests. 
A compaction thread flushes immutable memtables to StoCs and coordinates compaction of SSTables at StoCs. 
An xchg thread facilitates issuing requests to other StoCs and receiving their replies.
A reorganization (reorg) thread to balance load across dynamic ranges, Dranges.

\begin{figure}[!ht]
    \centering
    \includegraphics[width=0.8\linewidth]{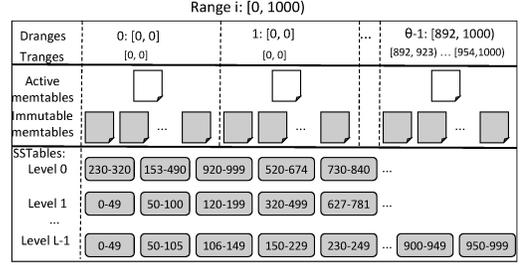}
    \caption{LTC constructs $\theta$ Dranges per range.}    \label{fig:ltc-Dranges}
\end{figure}

StoC has three types of threads.
A compaction thread compacts SSTables identified by an LTC and may write them to its local disk.
A storage thread reads and writes blocks from the local disk in response to an LTC request.
An xchg thread facilitates communication between these threads running on one StoC and those on other LTCs and StoCs. 
(StoCs communicate to perform compaction, see Section~\ref{sec:compaction}.)

The xchg thread continuously pulls on its QP for requests.
It backs-off exponentially when there is no request.  
This exponential back-off may be configured to strike a compromise between delays and the amount of utilized CPU.

\section{LSM-tree Component}\label{sec:ltc}
For each range, an LTC maintains an LSM-tree, multiple active memtables, immutable memtables, and SSTables at different levels. 
Nova-LSM further constructs $\theta$ dynamic ranges (Drange) per range, see Figure~\ref{fig:ltc-Dranges}. 
Dranges are transparent to an application. 

Each range maintains three invariants. 
First, key-value pairs in a memtable or a SSTable are sorted by key. 
Second, keys are sorted across SSTables at Level$_1$ and higher.
Third, key-value pairs are more up-to-date at lower levels, i.e., Level$_0$ has more up-to-date value of a key than Level$_i$ where i > 0.

\subsection{Dynamic Ranges, Dranges}\label{sec:dranges}
LTC uses a large number of memtables to saturate the bandwidth of StoCs. 
This raises several challenges. 
First, compacting SSTables at Level$_0$ into Level$_1$ becomes prohibitively long due to their large number as these Level$_0$ SSTables usually span the entire keyspace and must be compacted together.
Second, reads become more expensive since they need to search possibly all memtables and SSTables at Level$_0$. 
A skewed access pattern exacerbates these two challenges with a single key that is updated frequently.  
Different versions of this key may be scattered across all SSTables at Level$_0$.

To address these challenges, an LTC constructs Dranges with the objective to balance the load of writes across them evenly.  It assigns the same number of memtables to each Drange.  A write appends its key-value pair to the active memtable of the Drange that contains its key, preventing different versions of a single key from spanning all memtables.  SSTables at Level$_0$ written by different Dranges are mutually exclusive.  Thus, they may be compacted in parallel and independently.

An LTC monitors the write load of its $\theta$ Dranges.  It performs either a major or a minor reorganization to distribute its write load evenly across them.  Minor reorganizations are realized by constructing tiny ranges, Tranges, that are shuffled across Dranges.  Major reorganizations construct Dranges and Tranges based on the overall frequency distribution of writes.

A Drange is duplicated when it is a point and contains a very popular key.  For example, Figure~\ref{fig:ltc-Dranges} shows [0,0] is duplicated for two Dranges, 0 and 1, because its percentage of writes is twice the average.  This assigns twice as many memtables to Drange [0,0].  A write for key 0 may append to the active memtable of either duplicate Drange.  This reduces contention for synchronization primitives that implement thread-safe writes to memtables.  Moreover, the duplicated Dranges minimize the number of writes to StoC once a memtable is full as detailed in Section~\ref{sec:flush-memtable}.  
Below, we formalize Tranges, Dranges, minor and major reorganization.

\begin{definition}
A tiny dynamic range (Trange) is represented as [$TL_j$, $TU_j$). 
A Trange maintains a counter on the total number of writes that reference a key $K$ where $K < TU_j$ and $K \ge TL_j$. 
\end{definition}

\begin{definition}
A dynamic range (Drange) is represented as [$DL_i$, $DU_i$). A Drange contains a maximum of $\gamma$ Tranges where $DL_i = TL_{i,0}$, $DU_i = TU_{i,\gamma-1}$, and $\forall j\in[1,\gamma), TL_{i,j} = TU_{i,j-1}$. A Drange contains one active memtable and $\frac{\delta}{\theta}-1$ immutable memtables. Each Drange processes a similar rate of writes.
\end{definition}

\begin{definition}
A minor reorganization assigns Tranges of a hot Drange to its neighbors to balance load. 
\end{definition}

\begin{definition}
A major reorganization uses historical sampled data to reconstruct Tranges and their assigned Dranges.
\end{definition}

An LTC triggers a minor reorganization when a Drange receives a higher load than the average by a pre-specified threshold $\epsilon$, i.e., load > $\frac{1}{\theta}+\epsilon$. 
With a significant load imbalance where assigning Tranges to different Dranges does not balance the load, an LTC performs a major reorganization.  
It estimates the frequency distribution of the entire range by sampling writes from all memtables.  
Next, it constructs new Dranges and Tranges with the write load divided across them almost evenly.

An LTC may update the boundaries of a Drange$_i$ from $[DL_i, DU_i)$ to $[DL'_i, DU'_i)$ in different ways. 
% With a minor reorganization, there are different ways to implement assignment of a Trange from a hot Drange$_i$ to its colder neighbor Drange$_j$.
We consider two possibilities. 
With the first, these key-value pairs may be left behind in memtables of Drange$_i$.
% The first incurs the CPU overhead of copying and deleting data in memory.
This technique requires a get to search all memtables and overlapping SSTables at all levels since a higher level SSTable may contain the latest value of a key.
To elaborate, an older value of a key may reside in a memtable of Drange$_i$ while its latest value in a memtable of Drange$_j$ is flushed as a SSTable to Level$_0$. 
Subsequently, this SSTable may be compacted into higher levels earlier than the flushing of the memtable containing its older value (and belonging to Drange$_i$).

With the second, Nova-LSM associates a generation id with each memtable. 
This generation id is incremented after each reorganization. 
A reorganization marks the impacted active memtables as immutable, increments the generation id, and creates a new active memtable with the new generation id. 
When writing an immutable memtable as a SSTable, Nova-LSM ensures the memtables with older generation ids are flushed first. 
This allows a get to return immediately when it finds the value of its referenced key in a level.

One may come up with the best and worst case scenario for each of these techniques. 
The worst case scenario for the first technique is when a get references a key inserted least recently and appears in all levels.
In our experiment, we observed at most 40\% improvement with the second technique when gets are issued one at a time to the system. 
This improvement became insignificant with concurrent gets.

% Two possibilities are as follows.
% It may migrate the key-value pairs from original memtables of Dranges overlapping with range $[DL'_i, DU'_i)$ to Drange$_i$.
% Similarly, Drange$_i$ migrates the key-value pairs in its memtables that do not belong to range $[DL'_i, DU'_i)$ to the memtables of other Dranges that should contain those keys. 
% Second, 
% % Hence, a get that does not find its referenced key in either memtables or SSTables, must search all memtables and SSTables of higher levels. 
% We use lookup index to mitigate this overhead. 
% % The frequency of this overhead is minimized (maximized) when gets reference the latest (oldest) inserted keys.  
% This study focuses on the second approach and leaves the first approach and its comparison with the second to future work.
\subsubsection{Get}\label{sec:ltc:gets}
Each LTC maintains a lookup index to identify the memtable or the SSTable at Level$_0$ that contains the latest value of a key.
If a get finds its referenced key in the lookup index then it searches the identified memtable/SSTable for its value and returns it immediately.
%%%A get returns the value immediately if the referenced key is found in the lookup index. 
Otherwise, it searches overlapping SSTables at higher levels for the latest value and returns it. 
Each SSTable contains a bloom filter and LTC caches them in its memory.
A get skips a SSTable if the referenced key does not exist in its bloom filter.
In addition, a get may search SSTables at higher levels in parallel. 

Size of the lookup index is a function of the number of unique keys in memtables and SSTables at Level$_0$, $\ell_0$.  It is $\ell_0$*(the average key size + 4 bytes for memtable pointer + 8 bytes for Level$_0$ SSTable file number). 240 MB in experiments of Section~\ref{sec:eval}. 

The lookup index prevents a get from searching all memtables and SSTables at Level$_0$.  Its details are as follows.
A write that appends to a memtable $m$ updates the lookup index of the referenced key with $m$'s unique id $mid$. 
An LTC maintains an indirect mapping \textit{MIDToTable} from $mid$ to either a pointer to a memtable or the file number of a SSTable at Level$_0$. 
When a compaction thread converts an immutable memtable to a SSTable and flushes it to StoC, it atomically updates the entry of $mid$ in \textit{MIDToTable} to store the file number of the SSTable and marks the pointer to the memtable as invalid.

Once a SSTable at Level$_0$ is compacted into Level$_1$, its keys are enumerated.  For each key, if its entry in \textit{MIDToTable} identifies the SSTable at Level$_0$ then, the key is removed from the lookup index.

\subsubsection{Scan}\label{sec:ltc:scans}
An LTC maintains a range index to process a scan using only those memtables and Level$_0$  SSTables with a range overlapping the scan.
A scan must also search overlapping SSTables at higher levels since they may contain the latest values of the referenced keys. 

\begin{figure}[!ht]
    \centering
    \includegraphics[width=0.8\linewidth]{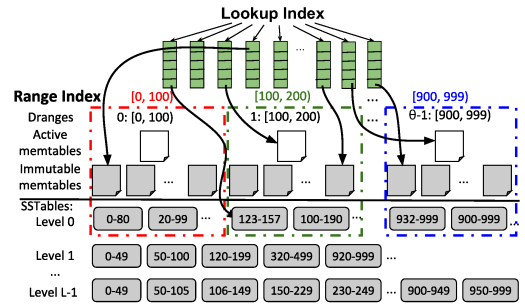}
    \caption{Lookup index and range index.}
    \label{fig:nova-lsm-index}
\end{figure}

Each element of the range index (a partition) corresponds to an interval, e.g., [0, 100) in Figure~\ref{fig:nova-lsm-index}.  It maintains a list of memtables and SSTables at Level$_0$ that contain keys within this range. 
These are in the form of pointers.  Size of required memory is the number of Dranges * (size of start interval key  + size of end interval key + pointers to memtables and Level$_0$ SSTable file numbers).  6 KB in experiments of Section~\ref{sec:eval}. 
%%%The range index exhibits minimal memory overhead as it only maintains the boundaries of partitions and a list of pointers to memtables and file numbers of Level$_0$ SSTables.

\begin{figure}[!ht]
    \centering
    \includegraphics[width=0.8\linewidth]{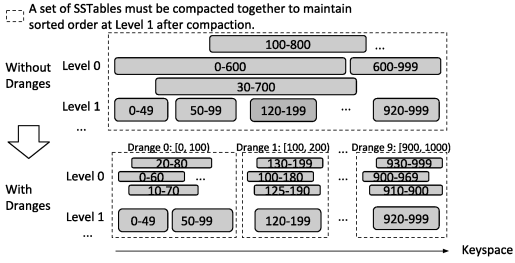}
    \caption{Dranges partition the keyspace for compaction.}
    \label{fig:compaction}
\end{figure}
\begin{figure}[!ht]
    \centering
    \includegraphics[width=0.8\linewidth]{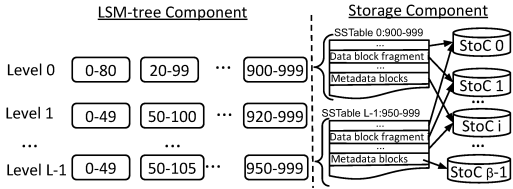}
    \caption{LTC scatters a SSTable across multiple StoCs.}
    \label{fig:ltc:scatter}
\end{figure}

A scan starts with a binary search of the range index to locate the partition that contains its start key. 
It then searches all memtables and SSTables at Level$_0$ in this partition. 
It also searches SSTables at higher levels. 
When it reaches the upper bound of the current range index partition, it seeks to the first key in the memtables/SSTables of the next range index partition and continues the search.

The range index is updated in three cases. 
First, when a new active memtable for a Drange or a new Level$_0$ SSTable is created, LTC appends it to all partitions of the index that overlap the range of this memtable/SSTable.
Second, a compaction thread removes flushed immutable memtables and deleted Level$_0$ SSTables from the range index.
Third, a Drange reorganization makes the range index more fine-grained by splitting its partitions. 
When a partition splits into two partitions, the new partitions inherit the memtables/SSTables from the original partition.
%%%Application may enforce a threshold on the maximum number of partitions.  No, because these are suppose to be transparent to the application.

\subsection{Flushing Immutable Memtables}\label{sec:flush-memtable}
Once a write exhausts the available space of a memtable, the processing LTC marks the memtable as immutable and assigns it to a compaction thread.  
This thread compacts the memtable by retaining only the latest value of a key and discarding its older values.  
If its number of unique keys is larger than a prespecified threshold, say 100, the compaction thread converts the immutable memtable into a SSTable and flushes it to StoC.  
Otherwise, it constructs a new memtable with the few unique keys, invokes LogC to create a new log file for the new memtable and a log record for each of its unique keys, and returns the new memtable as an immutable table to the Drange.  
When a Drange consists of multiple immutable memtables whose number of unique keys is lower than the prespecified threshold, the compaction thread combines them into one new memtable. Otherwise, it converts each immutable memtable to a SSTable and writes the SSTable to StoC.
With a skewed pattern of writes, this techniques reduces the amount of  data written to StoCs by 65\%.

\subsection{Parallel Compaction of SSTables at Level0}\label{sec:compaction}
Dranges divide the compaction task at Level$_0$ into $\theta$ smaller tasks that may proceed concurrently, see Figure~\ref{fig:compaction}. 
This minimizes write stalls when the total size of SSTables at Level$_0$ exceeds a certain threshold.
Without Dranges, in the worst-case scenario, a Level$_0$ SSTable may consist of keys that constitute a range overlapping all ranges of SSTables at Levels 0 and 1.  
Compacting this SSTable requires reading and writing a large volume of data from many SSTables. 
This worst-case scenario is divided $\theta$ folds by constructing $\theta$ non-overlapping Dranges.  
This enables $\theta$ concurrent compactions and utilizes resources that would otherwise sit idle. 
 
LTC employs a coordinator thread for compaction. 
This thread first picks Level$_i$ with the highest ratio of actual size to expected size.
It then computes a set of compaction jobs. 
Each compaction job contains a set of SSTables at Level$_i$ and their overlapping SSTables at Level$_{i+1}$.
SSTables in two different compaction jobs are non-overlapping and may proceed concurrently.

\noindent
\textbf{Offloading:}
When StoCs have sufficient processing capability, the coordinator thread offloads a compaction job to a StoC based on a policy.  In this study we assume round-robin.  A short-term research direction is to investigate other policies that minimize movement  of data using RDMA.
The StoC pre-fetches all SSTables in the compaction job into its memory. 
It then starts merging these SSTables into a new set of SSTables while respecting the boundaries of Dranges and the maximum SSTable size, e.g., 16 MB. 
It may either write the new SSTables locally or to other StoCs.
When the compaction completes, the StoC returns the list of StoC files containing the new SSTables to the coordinator thread. 
The coordinator thread then deletes the original SSTables and updates MANIFEST file described in Section~\ref{sec:ltc:crash}.

\subsection{SSTable Placement}\label{sec:scatter}
An LTC may be configured to place a SSTable across $\rho$ StoCs with $\rho \in [1, \beta]$ where $\beta$ is the number of StoCs.  With $\rho=1$, a SSTable is assigned to one StoC.  With $\rho>1$, the blocks of a SSTable are partitioned across $\rho$ StoCs. 
%%%Nova-LSM provides two placement strategies of a SSTable: 1) Assigning a SSTable to one StoC out of $\beta$ StoCs in its entirety, 2) Scattering its blocks across 
With large SSTables, partitioning reduces the time to write a SSTable as it utilizes the disk bandwidth of multiple StoCs. 

An LTC may adjust the value of $\rho$ for each SSTable using its size.
%%%Figure~\ref{fig:ltc:scatter} shows an LTC scatters a SSTable across a subset of StoCs with different SSTables scattered across all StoCs. 
For example, Figure~\ref{fig:ltc:scatter} shows SSTable 0:900-999 is scattered across 3 StoCs \{0, 1, i\} while SSTable L-1:950-999 is scattered across 4 StoC \{0, 1, i, $\beta$-1\}.  The first SSTable is smaller due to a skewed access pattern that resulted in fewer unique keys after compaction.
Hence, it is partitioned across fewer StoCs.

When LTC creates a SSTable, it first decides the value of $\rho$ and identifies the $\rho$ StoCs based on a policy such as 
%%A small $\rho$ results in larger writes using fewer StoCs. A large $\rho$ results in smaller writes using many StoCs. 
%%%Example policies to identify $\rho$ StoCs include 
random and power-of-d~\cite{power-of-two}.  
With random, an LTC selects $\rho$ StoCs from $\beta$ randomly.
With power-of-d~\cite{power-of-two}, LTC selects $\rho$ StoCs with the shortest disk queue out of $d$ randomly selected StoCs, d = $2 \times \rho$.

An LTC partitions a SSTable into $\rho$ fragments and identifies a StoC for each fragment.
It writes all fragments to $\rho$ StoCs in parallel. 
Finally, it converts its index block to StoC block handles and writes the metadata blocks (index and bloom filter blocks) to a StoC. 

\subsubsection{Availability}
Scattering a SSTable across multiple StoCs impacts its availability. 
If a SSTable is scattered across all $\beta$ StoCs, it becomes unavailable once a StoC fails.  One may scatter a SSTable across fewer StoCs and use either replication, a parity-based technique, or a hybrid of these technique to provide data availability.  %%%With replication, $R$ copies of a fragment of a SSTable fragment are constructed and stored on different StoCs.  With a parity-based technique, a parity block is computed using the fragments of a SSTable.  

Nova-LSM implements Hybrid, a technique that combines parity-based and replication technique as follows.  It constructs a parity block for the data fragments and replicates the metadata block due to its small size (in the order of a few hundred KBs).  This enables a read to fetch the metadata blocks from any replica using power-of-d.  

Fragments, their replicas, and parity blocks are assigned to different StoCs.  Their placements information are registered with the metadata block, see Step 5 of Figure~\ref{fig:flush_sstable}.

Hybrid does not incur the overhead of RAID to update a block since SSTables are immutable. 
Thus, the parity blocks are not read during normal mode of operation. 
In failed mode, the redundant data is read to recover content of a failed StoC.

We model the availability of data using the analytical models of~\cite{raid}.  
Assuming the Mean Time To Failure (MTTF) of a StoC is 4.3 months and repair time is one 1 hour~\cite{google-avail}, Table~\ref{tab:mttf} shows the MTTF of a SSTable and the storage layer consisting of 10 StoCs.  While MTTF of a SSTable ($MTTF_{SSTable}$) depends on the value of $\rho$, MTTF of the storage layer ($MTTF_{storage}$) is independent as we assume blocks of SSTables are scattered across all StoCs.  With no parity and no replication, $MTTF_{storage}$ is only 13 days.  

By constructing one replica for a SSTable and its fragments, $MTTF_{storage}$ is enhanced to 55.4 years and $MTTF_{SSTable}$ is improved to 554 years.  However, this technique has 100\% space overhead. 

A parity based technique has a lower space overhead than replication.  Its precise overhead depends on the value of $\rho$, see Table~\ref{tab:mttf}. 
% With $\rho$=1, we assume it constructs two replicas of a SSTable.  Higher values of $\rho$ reduce the space overhead.  They also reduce the availability of the system.  With $\rho$=3, a modest 33\% increase in storage space improves MTTF to tens of years. 
% Section~\ref{sec:eval} quantifies the impact of $\rho$ on system performance.  Using power-of-d, the performance with $\rho$=3 is almost the same as partitioning a SSTable across all StoCs, $\rho$=$\beta$.  See Table~\ref{tab:scatter}.
Ideally, StoCs should be deployed across different racks in a data center.  This preserves availability of data in the presence of rack failures.

%%%With $\rho=1$ and $R=2$, Mean Time To Failure of a SSTable is 554 years $MTTF_{SSTable}=554$ years, $MTTF_{storage}= 55.4$ years, and its disk space overhead is 100\%. 

%%%Assuming independent StoC failures and a StoC fails with a probability $p$. 
%%%The probability distribution of StoC failures follow the Binomial distribution. 
%%%The probability of $k$ failures $P(k)$ is ${\beta \choose k}p^k(1-p)^{\beta-k}$. 
%%%A SSTable is available with $k$ failures only if none of its $\rho$ fragments are placed on the $k$ failed disks. 
%%%We denote this probability using $Q(\rho,k)$. 
%%%\begin{equation}
%%%Q(\rho,k)= 
%%%\begin{cases}
%%%\frac{{\beta-k \choose \rho}}{{\beta \choose \rho}} & \text{if } \rho\leq \beta-k \\
%%%0 & \text{otherwise}
%%%\end{cases}
%%%\end{equation}
%%%The availability of a SSTable is 
%%%\begin{equation}
%%%\sum_{k=0}^{\beta}P(k)*Q(\rho,k)
%%\end{equation}

%%LTC computes a parity block using the SSTable fragments and meta blocks.  It then writes the parity block to another StoC. 

%%%Table~\ref{tab:mttf} shows the mean time to failure (MTTF) of a SSTable $MTTE_{SSTable}$, MTTF of the storage layer $MTTF_{storage}$, and disk space overhead as a function of $\rho$. 
%%%The reported numbers for RAID5 use the formula presented in~\cite{raid}. 
%%%We set MTTF of a StoC to 4.3 months~\cite{google-avail} and mean time to repair (MTTR) as 1 hour~\cite{raid}. 

\begin{table}
\small
\centering
\caption{Impact of $\rho$ on MTTF of a SSTable/Storage layer. %%%(*With $\rho=1$, use of parity constructs an additional replica.)
}
\begin{tabular}{ccccccc}
\hline
\multirow{2}{*}{$\rho$} & \multicolumn{2}{c}{$MTTF_{SSTable}$} & \multicolumn{2}{c}{$MTTF_{storage}$} & \multicolumn{2}{c}{Space overhead}\\\cline{2-7}
 & R=1 & Parity & R=1 & Parity & R=1 & Parity  \\\hline
1  & 4.3 Months & 554 Yrs & 13 Days  & 54 Yrs & 0\% & 100\% \\\hline
3  & 1.4 Months & 91 Yrs & 13 Days  & 30 Yrs  & 0\% & 33\% \\\hline
5  & 26 Days & 36 Yrs   & 13 Days  & 18.5 Yrs  & 0\% & 20\% \\\hline
% 10  & 310 & - & 310  & - & 1 & - \\\hline
\end{tabular}
\label{tab:mttf}
\end{table}

\subsection{Crash Recovery}\label{sec:ltc:crash}
Nova-LSM is built on top of LevelDB and reuses its well-tested crash recovery mechanism for SSTables and MANIFEST file. 
A MANIFEST file contains metadata of an LSM-tree, e.g., SSTable file numbers and their levels. 
Each application's specified range has its own MANIFEST file and is persisted at a StoC. 
Nova-LSM simply adds more metadata to the MANIFEST file. 
It extends the metadata of a SSTable to include the list of StoC file ids that store its meta and data blocks. 
It also appends the Dranges and Tranges to the MANIFEST file. 

When an LTC fails, the coordinator assigns its ranges to other LTCs.
With $\eta$ LTCs, it may scatter its ranges across $\eta-1$ LTCs.
This enables recovery of the different ranges in parallel. 
An LTC rebuilds the LSM-tree of a failed range using its MANIFEST file. 
Its LogC queries the StoCs for log files and uses RDMA READ to fetch their log records. 
The LTC then rebuilds the memtables using the log records, populating the lookup index. 
It rebuilds the range index using recovered Dranges and the boundaries of the memtables and Level$_0$ SSTables. 
% When a get observes a miss in the lookup index, it uses range index to search SSTables at Level$_0$ of the range index partition that contains the referenced key. 
% It also searches overlapping SSTables at all the other levels. 
% It populates the lookup index if the latest value is in a memtable or a SSTable at Level$_0$ and the key does not exist in the lookup index. 

\section{Logging Component}\label{sec:logc}
LogC constructs a log file for each memtable and generates a log record prior to writing to the memtable. 
LogC approximates the size of a log file to be the same as the memtable size.  
A log record is self-contained and is in the form of (log record size, memtable id, key size, key, value size, value, sequence number).  

The log file may be either in memory (availability) or persistent (durability). 
It may be configured to be persistent while its most recent log records are maintained in memory.  
This reduces mean-time-to-recovery of an LTC, enhancing availability while implementing durability.  
An in-memory log file may be replicated to enhance its availability.  
If all in-memory replicas fail, there is data loss. 
Next section describes our implementation of these alternatives.

\section{Storage Component}\label{sec:stoc}
A StoC implements in-memory and persistent StoC files for its clients, LTC and LogC. 
% We describe the implementation of each file type in turn. 

\subsection{In-memory StoC Files}
An in-memory StoC file consists of a set of contiguous memory regions. 
A StoC client appends blocks to the last memory region using RDMA WRITE. 
When the last memory region is full, it becomes immutable and the StoC creates a new memory region. 
A StoC client uses RDMA READ to fetch blocks. 

When a StoC client opens a file, the StoC first allocates a contiguous memory region in its local memory and returns the StoC file handle to the client. 
The file handle contains the StoC file id and the memory offset of the region. 
The client caches the file handle. 

A StoC client uses RDMA WRITE to append a block. 
It writes the block to the current memory offset at the StoC and advances the offset by the size of the block. 
If the block size is larger than the available memory, it requests a new memory region from the StoC and resumes the append.

We configure the size of a memory region to be the same size as a memtable to maximize performance. 
When LogC creates an in-memory StoC file for availability, only open and delete involve the CPU of the StoC. 
Appending a log record requires one RDMA WRITE. 
Fetching all of its log records requires one RDMA READ.
Both bypass StoC's CPU.

\subsection{Persistent StoC Files}
% A persistent StoC may store the data using alternative approaches that result in varying performance and price points, e.g., SSD, HDD or a hierarchical write-back. 
% With write-back, StoC first writes blocks to SSD to provide high performance. 
% It destages the blocks to HDD asynchronously. 

% Two alternatives for an StoC to materialize a persistent StoC file are 1) One Linux file per StoC file, 2) One Linux file shared across multiple StoC files.
% Sharing one Linux file provides higher performance by batching multiple blocks from different StoC files. 
% It incurs overhead for maintaining the mapping between StoC block handle and block handle in Linux files. 
% Also, fetching all blocks in an StoC file has a higher overhead since it requires scanning many Linux files to retrieve its blocks. 
% It also imposes a higher storage cost if deleting blocks of an StoC file is performed lazily, e.g., a Linux file is deleted when all its contained StoC blocks are obsolete. 
% However, we observe a marginal performance improvement with sharing during our evaluation. 
% Hence, we detail the design of one Linux file per StoC file here.

A StoC maintains a file buffer shared across all persistent StoC files.
When a StoC client appends a block, it first writes the block to the file buffer using RDMA WRITE. 
When a StoC client reads a block, the StoC fetches the block into the file buffer and writes the block to the StoC client's memory using RDMA WRITE. 

\begin{figure}[!ht]
    \centering
    \includegraphics[width=0.8\linewidth]{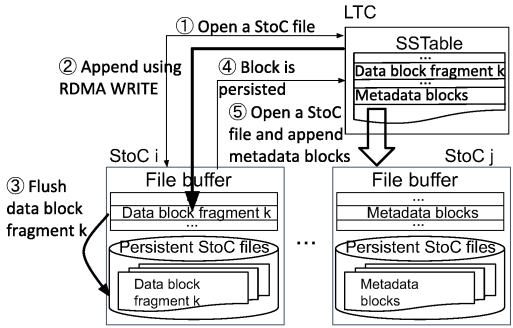}
    \caption{The workflow between one LTC and multiple StoCs to scatter blocks of a SSTable.}
    \label{fig:flush_sstable}
\end{figure}

Figure~\ref{fig:flush_sstable} shows the workflow of an LTC appending a data block fragment to a StoC file. 
Its StoC client first sends a request to open a new StoC file~\circled{1}. 
The StoC allocates a memory region in its file buffer. 
The size of the memory region is the same as the block size.
It also creates a unique id for this memory region and returns its memory offset to the client along with this unique id.
The client then writes the block to this memory region using RDMA WRITE. 
It sets this unique id in the immediate data~\circled{2}. 
The StoC is notified once the RDMA WRITE completes. 
It then flushes the written block to disk~\circled{3} and sends an acknowledgment to the client~\circled{4}. 
Lastly, it writes the metadata blocks to a StoC following the same procedure~\circled{5}.

\section{Implementation}\label{sec:impl}
We implemented Nova-LSM by extending LevelDB with 20,000+ lines of C++ code that implements components of Sections 3 to 6, an LTC client component, the networking layer for an LTC to communicate with its clients, and management of memory used for RDMA’s READ and WRITE verbs.  With the latter, requests for different sized memory allocate and free memory from a fixed preallocated amount of memory.  We use memcached’s slab allocator~\cite{memcached} to manage this memory.

%%%We added a networking layer on top of LTC to process client requests issued over the IP Network.  We ported the slab allocator from memcached~\cite{memcached} to manage the RDMA memory and the file buffer.  We implemented LogC to replicate log records across StoCs over RDMA to provide high availability. 

% Each RDMA QP has a circular buffer to post requests.  
% It posts 32 receive events to accept requests from the connected QP on the other server.
% An RDMA thread polls the completion queues of its QPs continuously to maximize performance. 
% To reduce CPU usage when the system is idle, it backs off when there are no completed events.

% An StoC client attaches a monotonically increasing request id to every request issued to an StoC.
% The StoC also piggybacks this request id in its response to the StoC client. 
% The request id is set in the immediate data of an RDMA request. 
% In this way, an StoC client knows the associated request when it receives a response. 

\section{Evaluation}\label{sec:eval}
This section evaluates the performance of Nova-LSM.  We start by quantifying the performance tradeoffs associated with different configuration settings of Nova-LSM.  Next, we compare Nova-LSM with LevelDB and RocksDB.  Main lessons are as follows: 
\begin{itemize}[leftmargin=*]

\item When comparing performance of Nova-LSM with and without  Dranges, Dranges enhance throughput from a factor of 3 to 26.  (Section~\ref{sec:eval:Dranges})

\item With CPU intensive workloads, one may scale LTCs to enhance throughput.  With disk I/O intensive workloads, one may scale StoCs to enhance throughput. (Sections~\ref{sec:eval:vertical-scale-memory} and~\ref{sec:eval:horizontal-scale})

\item With a skewed workload that requires many requests to reference ranges assigned to a single LTC, one may re-assign ranges of the bottleneck LTC to other LTCs to balance load.  This enhances performance from 60\% to more than 4 folds depending on the workload.  (Section~\ref{sec:eval:ltc-load-balancing})

\item Nova-LSM outperforms both LevelDB and RocksDB by more than an order of magnitude with all workloads that use a skewed pattern of access to data.  It provides comparable performance to LevelDB and RocksDB for most workloads with a uniform access pattern.
\item Nova-LSM is inferior with CPU intensive workloads and a uniform access pattern due to its higher CPU usage.  With a single node, the CPU overhead is due to maintaining the index.  With more than one nodes, the CPU overhead also includes xchg threads pulling for requests. (Section~\ref{sec:eval:leveldb})

\end{itemize}

\subsection{Experiment Setup}
We conduct our evaluation on the CloudLab APT cluster of c6220 nodes~\cite{cloudlab}. 
Each node has 64 GB of memory, two Xeon E5-2650v2 8-core CPUs, and one 1 TB hard disk. 
Hyperthreading results in a total of 32 virtual cores. 
All nodes are connected using Mellanox SX6036G Infiniband switches. 
Each server is configured with one Mellanox FDR CX3 Single port mezz card that provides a maximum bandwidth of 56 Gbps and a 10 Gbps Ethernet NIC. 

Our evaluation uses the YCSB benchmark~\cite{ycsb} with different sized databases: 10 GB (10 million records), 100 GB (100 million records), 1 TB (1 billion records), and 2 TB (2 billion records).
Each record is 1 KB. 
We focus on three workloads shown in Table~\ref{tab:workloads}. 
A get request references a single key and fetches its 1 KB value. 
A write request puts a key-value pair of 1 KB. 
A scan retrieves 10 records starting from a given key. 
If a scan spans two ranges, we process it in a read committed~\cite{ansi-isolation-levels} manner with puts. 
YCSB uses Zipfian distribution to model a skewed access pattern.  
We use the default Zipfian constant 0.99, resulting in 85\% of requests to reference 10\% of keys. 
With Uniform distribution, a YCSB client references each key with almost the same probability. 

\begin{table}
\small
\centering
\caption{Workloads.}
\begin{tabular}{ll} \hline
Workload & Actions \\ \hline
RW50 & 50\% Read, 50\% Write. \\ \hline
SW50 & 50\% Scan, 50\% Write. \\ \hline
W100 & 100\% Write. \\ \hline
\end{tabular}
\label{tab:workloads}
\end{table}

A total of 60 YCSB clients running on 12 servers, 5 YCSB clients per server, generate a heavy system load.  
Each client uses 512 threads to issue requests. 
A thread issues requests to LTCs using the 10 Gbps NIC.
We use a high number of worker/compaction threads to maximize throughput by utilizing resources that would otherwise sit idle.
An LTC has 512 worker threads to process client requests and 128 threads for compaction.
An LTC/StoC has 32 xchg threads for processing requests from other LTCs and StoCs, 16 are dedicated for performing compactions~\cite{rdma-eval,rdma-guideline}. 
A StoC is configured with a total of 256 threads, 128 are dedicated for compactions. 
Logging is disabled by default. 
When logging is enabled, LogC replicates a log record 3 times across 3 different StoCs.

% With both LevelDB and NOVA-LSM, an application may partition data into $\omega$ ranges and assign them to different instances of LevelDB and LTC.
% This partitioning information is provided to each YCSB client to direct a request to the appropriate instance.
% With both systems, we set memtable size and SSTable size to 16 MB, $\tau=16$.
% Moreover, both are configured to maintain a bloom filter with each SSTable that has 1\% false positive rate. 

% With $\alpha$ active memtables, Nova-LSM constructs $\alpha$ Dranges dynamically at runtime, $\theta=\alpha$. 
% We use 10 tiny ranges per Drange, $\gamma=10$.

% We configure LevelDB and Nova-LSM with 6 levels. 
% The expected size of Level$_{0}$ is 4 GB. 
% The expected size of Level$_{i}$ is 3.2 times larger than Level$_{i-1}$. 
% The maximum size of Level$_0$ is 10 GB. 
% The maximum Level$_0$ size and expected size of a level is evenly divided across $\omega$ ranges. 
% For example, with $\omega=64$ ranges, the maximum number of Level$_0$ SSTables is 10 per range and compacting SSTables at Level$_0$ starts when the number of Level$_0$ SSTables is 4. 
% % We compact the level with the highest ratio of actual size to expected size. 

% We first load all records into the database and perform a warmup for 10 minutes by generating puts. 
% We terminate the warmup when there are no SSTables at Level$_0$ and the actual size of each level is below its expected size. 
% Next, we run an experiment for a specific workload (see Table~\ref{tab:workloads}) for 20 minutes.
% We report the aggregated throughput across all YCSB clients. 

\subsection{Nova-LSM and Its Configuration Settings}
This section quantifies the performance tradeoffs associated with alternative settings of Nova-LSM.  
We use a 10 GB database and one range per LTC, $\omega=1$. 
Each server hosts either one LTC or one StoC. 
Section~\ref{sec:eval:Dranges} evaluates Dranges. 
Section~\ref{sec:eval:logging} evaluates the performance impact of logging. 
One may scale each component of NOVA-LSM either vertically or horizontally. 
Its configuration settings may scale (1) the amount of memory an LTC assigns to a range, (2) the number of StoCs a SSTable is scattered across, (3) the number of StoCs used by different SSTables of a range, and (4) the number of LTCs used to process client requests. 
Items 1 is vertical scaling. 
The remaining three scale either LTCs or StoCs horizontally. 
We discuss them in turn in Section~\ref{sec:eval:vertical-scale-memory} and Section~\ref{sec:eval:horizontal-scale}, respectively.
Section~\ref{sec:eval:ltc-load-balancing} describes migration of ranges across LTCs to balance load.
We present results on the recovery duration in Section~\ref{sec:eval:recovery}. 

\subsubsection{Dynamic Ranges}\label{sec:eval:Dranges}
Nova-LSM constructs Dranges dynamically at runtime based on the pattern of access to the individual Dranges.
Its objective is to balance the write load across the Dranges. 
We quantify {\em load imbalance} as the standard deviation on the percentage of puts processed by different Dranges.
The ideal load imbalance is 0. 
We report this number for both Uniform and Zipfian.
The reported numbers for Uniform serve as the desired value of load imbalance for Zipfian.

With Uniform and $\theta=64$ Dranges, we observe a very small load imbalance, 2.86E-04$\pm$3.21E-05 (mean$\pm$standard deviation), across 5 runs. 
Nova-LSM performs only one major reorganization and no minor reorganizations.
With Zipfian, the load imbalance is 1.65E-03$\pm$4.42E-04. 
This is less than 6 times worse than Uniform. 
Nova-LSM performs one major reorganization and 14.2$\pm$14.1 minor reorganizations. 
There are 11 duplicated Dranges. 
The first Drange [0,0] is duplicated three times since key-0 is accessed the most frequently.
Similar results are observed as we vary the number of Dranges from 2 to 64.  
Duplication of Dranges containing one unique key is essential to minimize load imbalance with Zipfian. 

\begin{figure}
    \centering
    \includegraphics[width=\linewidth]{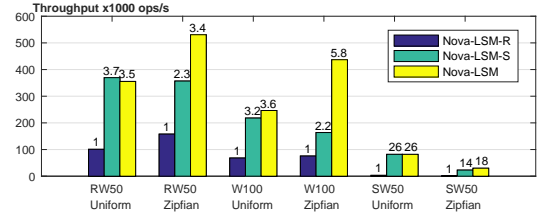}
    \caption{Throughput comparison of Nova-LSM and its variants with $\eta=1$, $\beta=10$, $\rho=1$, $\alpha=64$, and $\delta=256$.}
    \label{fig:eval:nova-lsm-sr-opt}
\end{figure}

Dranges facilitate concurrent compactions. 
We compare Nova-LSM with Nova-LSM-R and Nova-LSM-S, see Figure~\ref{fig:eval:nova-lsm-sr-opt}. 
Nova-LSM-R randomly selects an active memtable to process a put request. 
Nova-LSM-S uses the active memtable of the corresponding Drange to process a put request. 
Both do not prune memtables or merge immutable memtables into a new memtable for Dranges with fewer than 100 unique keys.

With RW50 and W100, Nova-LSM provides 3x to 6x higher throughput when compared with Nova-LSM-R as it compacts SSTables in parallel.
With Nova-LSM-R, a SSTable at Level$_0$ spans the entire keyspace and overlaps with all other SSTables at Level$_0$ and Level-1. 
It must first split these sparse SSTables at Level$_0$ into dense SSTables and then compact them in parallel. 
The throughput of Nova-LSM is comparable to Nova-LSM-S with Uniform. 
It is higher with Zipfian because compaction threads write fewer data to StoCs. 
With Zipfian, Nova-LSM constructs 11 Dranges that correspond to one unique key and a total of 26 Dranges that contain less than 100 unique keys.
With these Dranges, Nova-LSM merges their memtables into a new memtable instead of flushing them to StoCs, providing significant savings by avoiding disk reads and writes. 

With SW50, Nova-LSM outperforms Nova-LSM-R by 26x with Uniform and 18x with Zipfian. 
This is because keys are scattered across memtables and SSTables at Level$_0$ with Nova-LSM-R. 
With Uniform, a scan searches 64 memtables and 75 SSTables at Level$_0$ on average while Nova-LSM searches only 1 memtable and 2 SSTables.

\subsubsection{Impact of Skew}
\begin{figure}
    \centering
    \includegraphics[width=\linewidth]{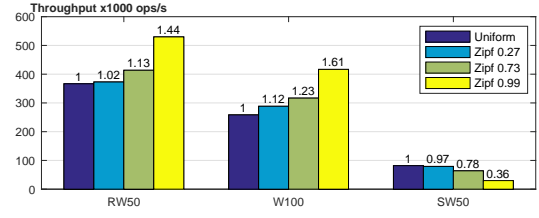}
    \caption{Impact of Skew with $\eta=1$, $\beta=10$, $\rho=1$, $\alpha=64$, and $\delta=256$.}
    \label{fig:eval:nova-lsm-skew}
\end{figure}

Figure~\ref{fig:eval:nova-lsm-skew} shows the throughput of Nova-LSM for different workloads as a function of skewness.  
The access pattern becomes more skewed as we increase the mean of the Zipfian distribution from 0.27 to 0.99.  
The number at top of each bar is the factor of improvement relative to Uniform.

The throughput of both RW50 and W100 increases with a higher degree of skew.  
It decreases with SW50.  
Consider each workload in turn.  
With RW50, Nova-LSM processes get requests using memory instead of SSTables by maintaining the most popular keys in the memtables.  
With W100, Nova-LSM compacts memtables of Dranges with fewer than 100 unique keys into a new memtable without flushing them to disk.   
The number of Dranges with fewer than 100 unique keys increases with a higher degree of skew.  
Hence, the percentage of experiment time attributed to write stalls decreases from 46\% with Uniform to 16\% with the most skewed access pattern. 

With SW50, the CPU is the limiting resource with all distributions including Uniform.  
A higher skew generates more versions of a hot key.  A scan must iterate these versions to find the next unique key that may reside in a different memtable or SSTable.  This increases CPU load with a higher degree of skew, decreasing the observed throughput.  

\subsubsection{Logging}\label{sec:eval:logging}
When the CPU of LTC is not utilized fully, logging imposes a negligible overhead on service time and throughput. 
We measure the service time of a put request from when a client issues a put request until it receives a request complete response.
Replicating a log record three times using RDMA observes only a 4\% overhead compared to when logging is disabled (0.51 ms versus 0.49 ms). 
When replicating log records using NIC, the service time increases to 1.07 ms (2.1x higher than RDMA) since it involves the CPU of StoCs. 

With replication only, logging uses CPU resources of LTCs and impacts throughput when CPU of one or more LTCs is fully utilized. 
As shown in Figure~\ref{fig:eval:nova-leveldb-10-servers}, logging has negligible overhead with Uniform since the CPU is not fully utilized. 
With Zipfian, it decreases the throughput by at most 33\% as the CPU utilization of the first LTC is higher than 90\%. 
Logging has a negligible impact on the CPU utilization of StoCs since LogC uses RDMA WRITE to replicate log records bypassing CPUs of StoCs. 

% \begin{table}
% \small
% \centering
% % put all parameters here. 
% \caption{Service time of a put request.}
% \begin{tabular}{ll} \hline
% Logging method & Service time (ms)  \\ \hline
% Disabled & 0.49 \\ \hline
% RDMA (3 replicas) & 0.51 \\ \hline
% NIC (3 replicas) & 1.07 \\ \hline
% Disk (1 replica) & 33.41 \\ \hline
% \end{tabular}
% \label{tab:eval:log-service-time}
% \end{table}

\begin{table}
\small
\centering
\caption{Throughput of W100 Uniform as a function of the memory size with $\eta=1$, $\beta=10$, and $\rho=1$.}
\begin{tabular}{lccc}
%%%\begin{tabular}{p{2cm}p{0.5cm}p{0.5cm}p{3cm}} 
\hline
Memory size & $\alpha$ & $\delta$ &  Throughput (ops/s) \\ \hline
32 MB&    1&    2&    8,924 \\ \hline
64 MB&    2&    4&    10,594 \\ \hline
128 MB&    4&    8&    36,095 \\ \hline
256 MB&    8&    16&    34,102 \\ \hline
512 MB&    16&    32&    53,258 \\ \hline
1 GB&    32&    64&    204,774 \\ \hline
2 GB&    64&    128&    245,753 \\ \hline
4 GB&    64&    256&    246,434 \\ \hline
\end{tabular}
\label{tab:eval:vertical-mem}
\end{table}
\subsubsection{Vertical Scalability}\label{sec:eval:vertical-scale-memory}
One may vertically scale resources assigned to an LTC and/or a StoC either in hardware or software.  
The amount of memory assigned to an LTC has a significant impact on its performance.  
(Challenge 1 of Section 1 demonstrated this by showing a 6 fold increase in throughput as we increased memory of an LTC from 32 MB to 4 GB.)
It is controlled using the following parameters:  number of ranges ($\omega$), number of memtables per range ($\delta$), number of active memtables per range ($\alpha$), and memtable size ($\tau$).  
Since $\omega$ ranges may be assigned to different LTCs, we focus on $\omega$=1 range and discuss the impact of $\delta$ and $\alpha$ on throughput of W100.  
We set the maximum degree of parallelism for compaction to be $\alpha$.
Experiments of this section use a configuration consisting of 1 LTC with 10 StoCs.  Each SSTable is assigned to 1 StoC ($\rho$=1).  
We use $\beta=10$ to avoid the disk from becoming the bottleneck. 
With $\beta=1$, the disk limits the performance and the LTC does not scale vertically as a function of additional memory.

Our experiments start with a total of 2 memtables ($\delta$=2), requiring 32 MB of memory, Memory size = $\delta \times 16$ MB.
1 memtable is active ($\alpha$=1),   
We double memory size by increasing values for both $\alpha$ and $\delta$ two-folds, see Table~\ref{tab:eval:vertical-mem}.  
With sufficient memory, the throughput scales super-linearly. 
For example, increasing memory from 512 MB to 1 GB increases the throughput almost 4 folds. 
This is because the duration of write stalls reduces from 65\% to 21\% of experiment time.
The throughput levels off beyond 2 GB as the bandwidth of StoCs becomes fully utilized.

\subsubsection{Horizontal Scalability}\label{sec:eval:horizontal-scale}
Horizontal scalability of Nova-LSM is impacted by the number of LTCs, StoCs, how a SSTable is scattered across StoCs and whether the system uses power-of-d.  
Once the CPU of an LTC or disk bandwidth of a StoC becomes fully utilized, it dictates the overall system performance and its scalability.  
Below, we describe these factors in turn.  

\begin{figure}
\begin{subfigure}[t]{0.47\columnwidth}
\centering
\includegraphics[width=\textwidth]{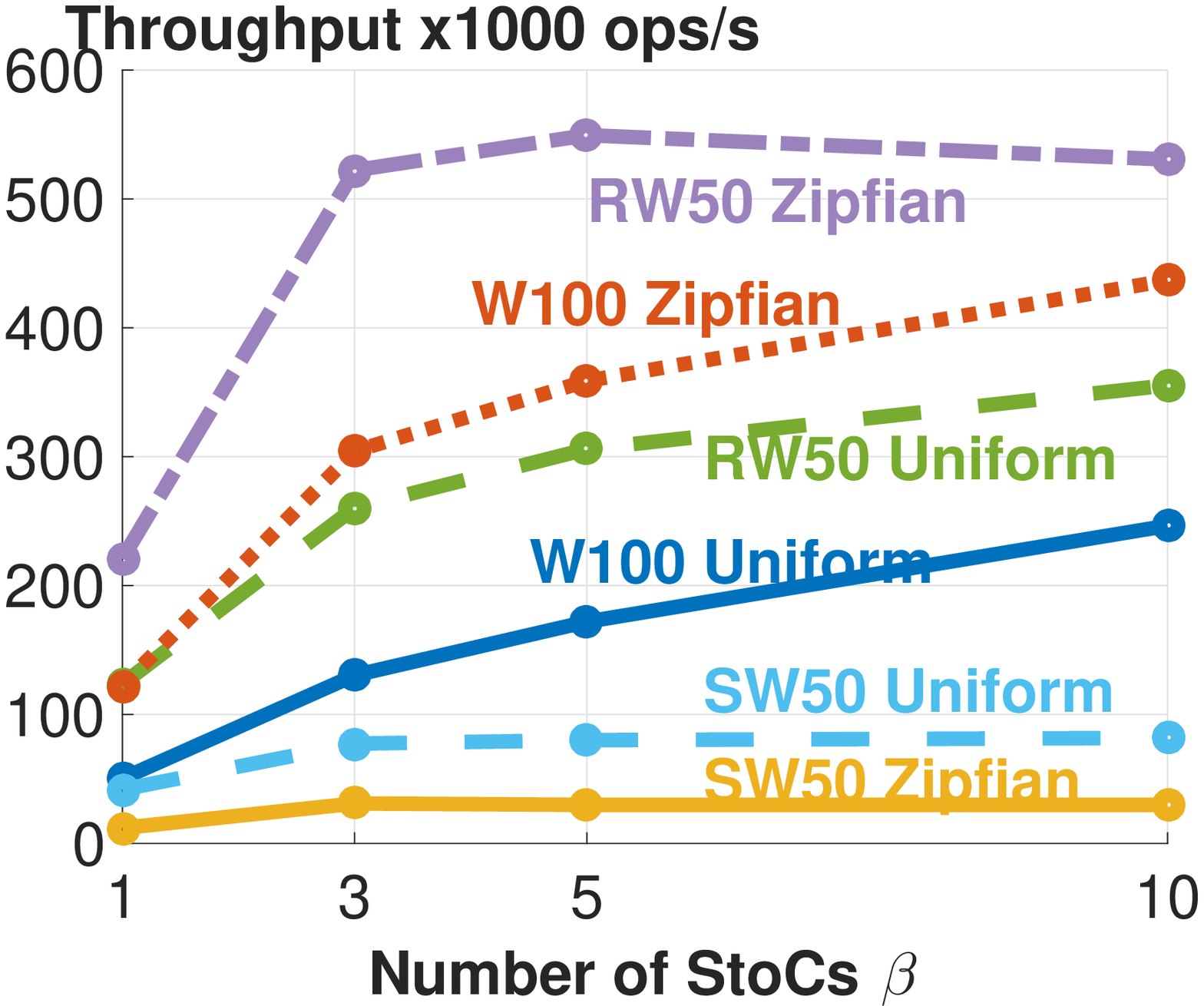}
\caption{Throughput.}
\label{fig:eval:thpt-stocs}
\end{subfigure}
\quad
\begin{subfigure}[t]{0.47\columnwidth}
\centering
\includegraphics[width=\textwidth]{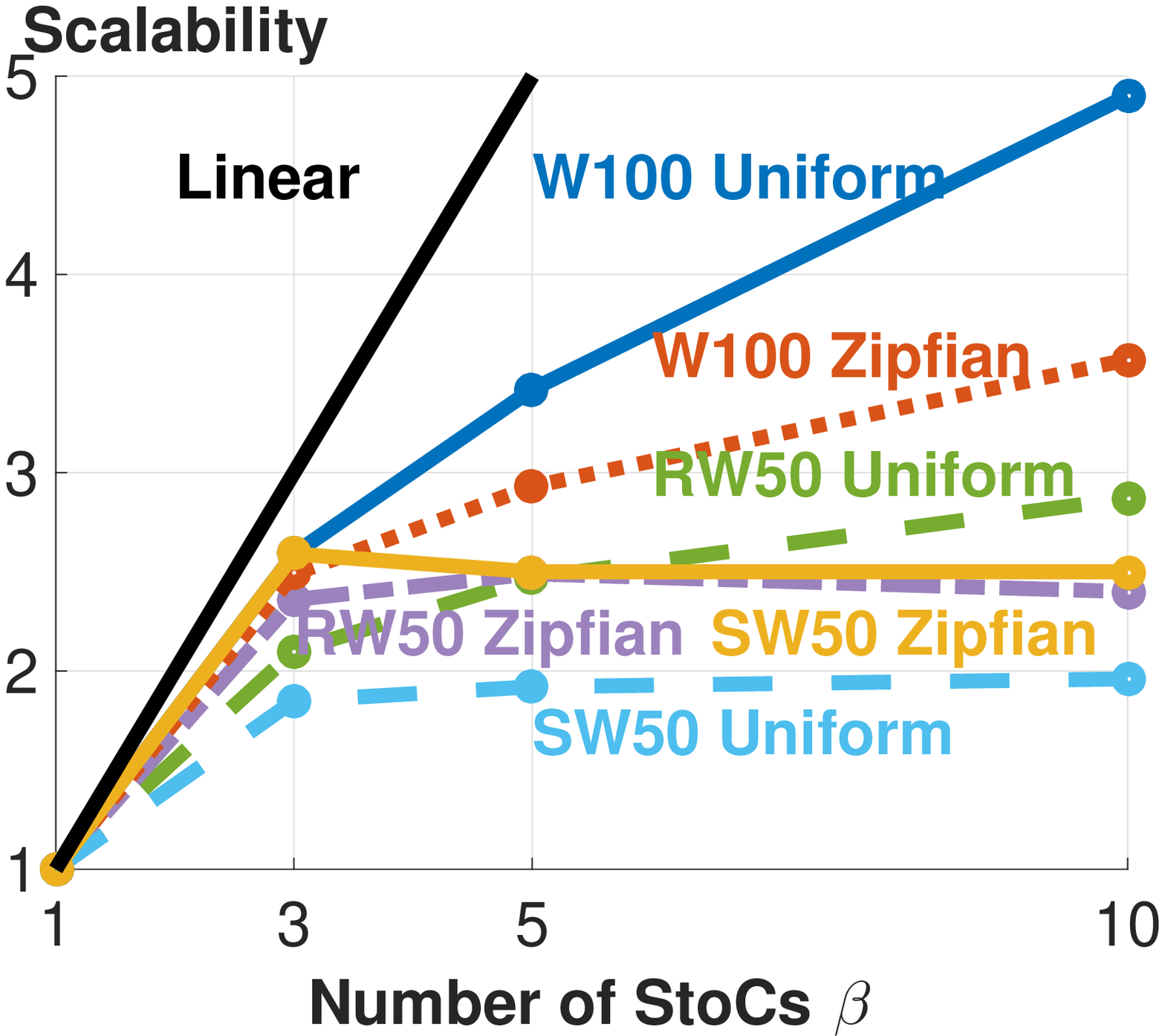}
\caption{Scalability.}
\label{fig:eval:scale-stocs}
\end{subfigure}
\caption{Throughput and scalability as a function of $\beta$ with $\eta=1$, $\rho=1$, $\alpha=64$, and $\delta=256$. }
\label{fig:eval:stocs}
\end{figure}

\noindent{\bf Scalability of StoCs:}  Figure~\ref{fig:eval:stocs} shows the throughput and horizontal scalability of a system consisting of one LTC as we increase the number of StoCs, $\beta$, from one to 10.  
LTC stores a SSTable in one StoC ($\rho=1$) using power-of-2.
The throughput scales as long as the CPU of the server hosting the single LTC does not become a bottleneck.  
Once the CPU becomes fully utilized, adding StoCs does not provide a performance benefit.  
With RW50 and SW50, this limit is reached with 5 StoCs.  
Hence, additional StoCs do not provide a benefit.  
Amongst the different workloads, W100 scales the best.  
With Zipfian, the CPU of the single LTC limits its scalability.
With Uniform, the write stalls due to the maximum Level$_0$ size limits the throughput\footnote{The throughput is 319K when the maximum Level$_0$ size is 20 GB. It increases further to 351K when the maximum Level$_0$ size increases to 30 GB.}. 
The write stall duration decreases with more StoCs. 
However, it is not a linear function of the number of StoCs. 
The percentage of experiment time spent on write stalls decreases from 82\% with one StoC to 60\% with three and 46\% with 10 StoCs. 
This is because LTC schedules compactions across StoCs in a round-robin manner, resulting in one or more StoCs to become fully utilized. 
We will consider policies other than round-robin as future work. 

\begin{table}
\small
\centering
\caption{Throughput of W100 Uniform as a function of $\rho$ with $\eta=1$, $\beta=10$, $\alpha=1$, $\delta=2$.}
\begin{tabular}{ccc}
%%%\begin{tabular}{p{1cm}p{2cm}p{3cm}} 
\hline
$\rho$ & Random & Power-of-$d$\\ \hline
% 1 & 35,511 &	52,865 \\ \hline
% 3 & 44,180 &	58,942 \\ \hline
% 10 & 58,098 &	58,877 \\ \hline
1 & 27,593 &     42,659  \\ \hline
3 & 42,016     &50,433  \\ \hline
10 & 52,590 &    52,477  \\ \hline
\end{tabular}
\label{tab:eval:scatter:2memtable}
\end{table}
\noindent{\bf Power-of-d and degree of partitioning $\rho$:} One may configure NOVA-LSM to write a SSTable to a subset ($\rho$) of StoCs instead of all $\beta$ StoCs.
The value of $\rho$ has a significant impact on the performance of a deployment.  
With a limited amount of memory (32 MB), using all 10 disks ($\rho$=10) provides a throughput that is almost twice higher than with one disk ($\rho$=1), see Table~\ref{tab:eval:scatter:2memtable} and the column labeled Random.  
This is because when both memtables are full, a write must wait until one of the memtables is flushed to StoC.  
The time to perform this write is longer when using one disk compared with 10 disks, allowing $\rho$=10 to provide a significant performance benefit.

Table~\ref{tab:eval:scatter:2memtable} shows power-of-2 provides 54\% higher throughput than random when $\rho=1$. 
It minimizes the queuing delays attributed to writings of SSTables colliding on the same StoC. 
When $\rho=3$, power-of-6 provides a comparable throughput as $\rho=10$. 
A smaller $\rho$ results in larger sequential writes that enhances the performance of hard disks. 
This is not highlighted in Table~\ref{tab:eval:scatter:2memtable}.  
As an example, with 5 LTCs and W100 using Uniform, 
when $\rho$=3 is compared with $\rho$=10, its observed throughput is 17\% higher.

Figure~\ref{fig:intro-performance} in Section~\ref{sec:intro} presents results for two configurations of Nova-LSM:  shared-nothing and shared-disk.  Both use the same hardware platform consisting of 10 nodes.  Each node has its own local disk drives.  With the shared-nothing, an LTC stores its SSTables on its local hard disk drive.  With the shared-disk, an LTC partitions a SSTable across $\rho$=3 StoCs (using power-of-d) out of $\beta$=10 StoCs. In essence, the shared-nothing does not use the RDMA while shared-disk uses the RDMA to enable LTCs to share disk bandwidth.  

When compared with one another, the shared-disk provides the highest performance gains with YCSB’s default Zipfian mean setting, 0.99.  It utilizes the bandwidth of all 10 disks while the shared-nothing causes the disk bandwidth of the node containing the most popular data to become the bottleneck.  The degree of skew impacts the reported factor of improvements with the shared-disk.  For example, with a less skewed access pattern, say Zipfian mean of 0.73 that directs 53\% of requests to 10\% of the keys, the factor of improvement is reduced to 3.8, 3.7, and 5.7 with RW50, W100, and SW50 workloads, respectively.  The reduced load of the bottleneck server in the shared-nothing architecture minimizes the impact of sharing disk bandwidth and the factor of improvement.

\begin{figure}
\begin{subfigure}[t]{0.47\columnwidth}
\centering
\includegraphics[width=\textwidth]{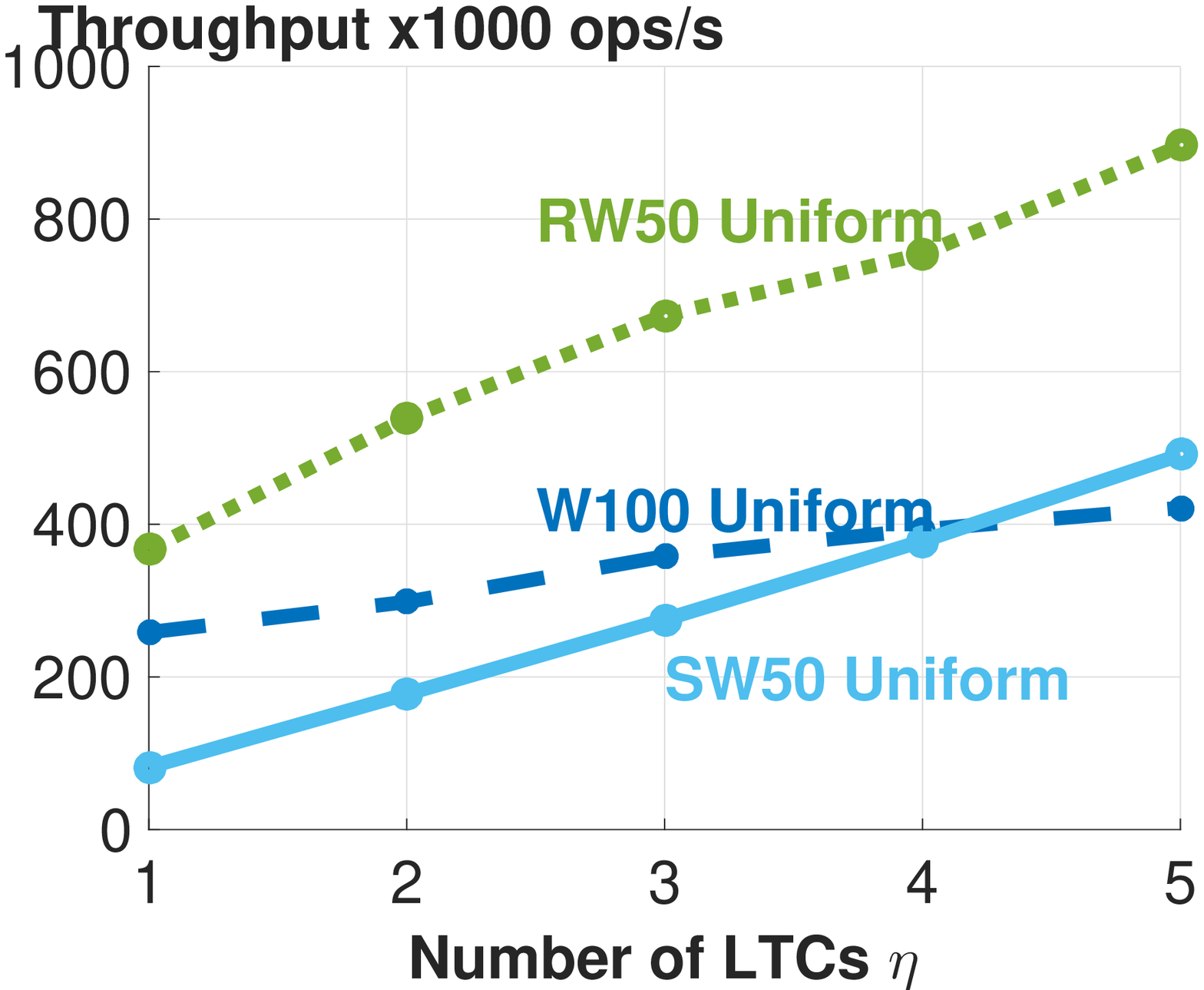}
\caption{Throughput.}
\label{fig:eval:thpt-ltcs}
\end{subfigure}
\quad
\begin{subfigure}[t]{0.47\columnwidth}
\centering
\includegraphics[width=\textwidth]{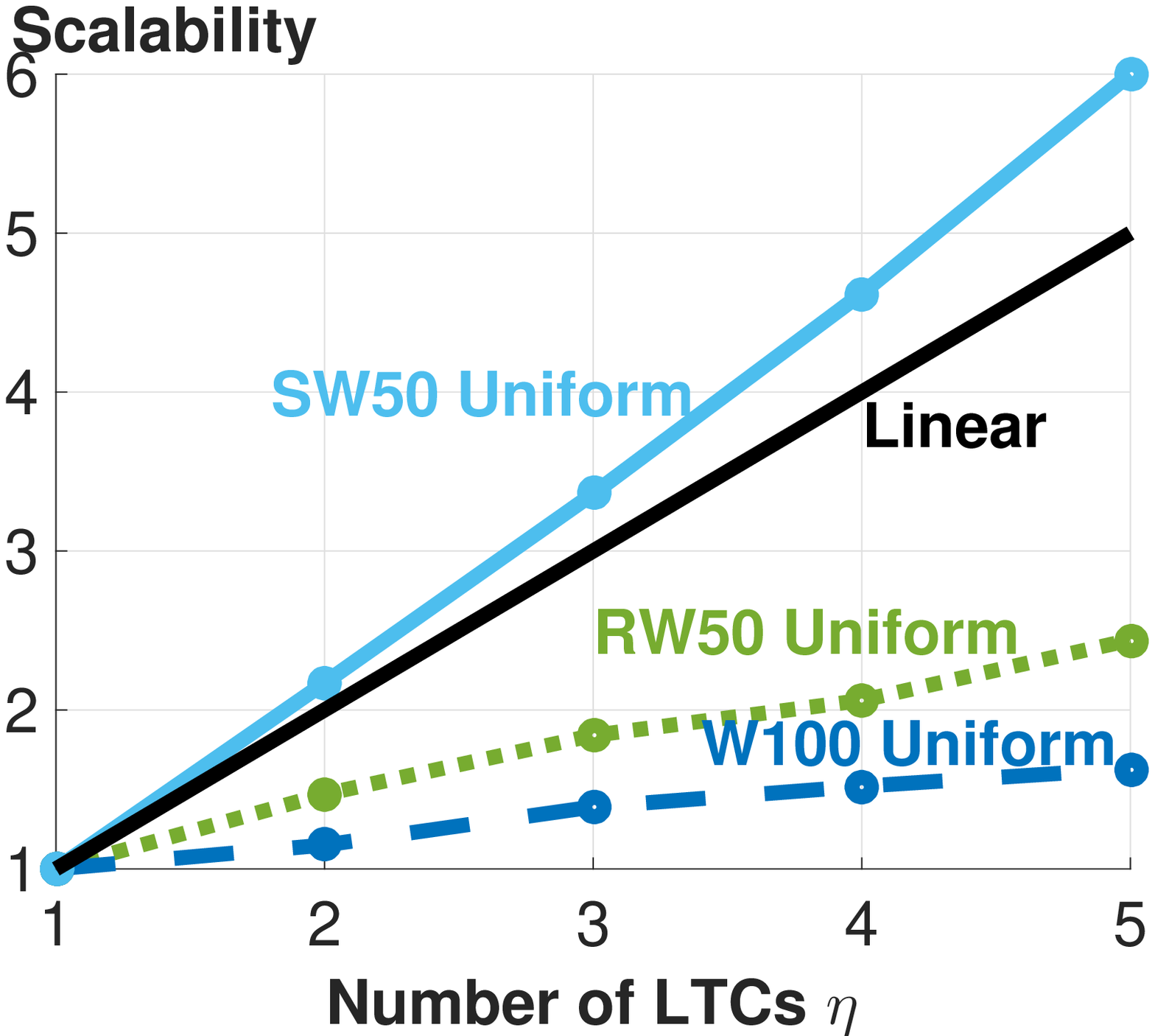}
\caption{Scalability.}
\label{fig:eval:scale-ltcs}
\end{subfigure}
\caption{Throughput and scalability as a function of $\eta$ with $\beta=10$, $\rho=3$, $\alpha=64$, and $\delta=256$.}
\label{fig:eval:scale-ltc}
\end{figure}

\noindent{\bf Scalability of LTCs:} One may increase the number of LTCs in a configuration to prevent its CPU from becoming the bottleneck. 
Figure~\ref{fig:eval:scale-ltc} shows system throughput as we increase the number of LTCs from 1 to 5 with Uniform. 
The system is configured with a total of 256 memtables of which 64 are active and 10 StoCs.
Each SSTable is partitioned across 3 StoCs, $\rho$=3.
All reported results are obtained using power-of-6.

The SW50 workload scales super-linearly.  The database size is 10 GB in this experiment.  With 1 LTC and 4 GB of memtables, the database does not fit in memory.  With 5 LTCs, the database fits in the memtables of the LTCs, reducing the number of accesses to StoCs for processing scans.  It is important to note that the CPUs of the nodes hosting LTCs remain fully utilized.

The RW50 workload scales sub-linearly because, while the CPU of 1 LTC is fully utilized, it is not fully utilized with 2 or more LTCs.  Moreover, the percentage of time the experiment spends in write stall increases from 4\% with 1 LTC to 13\% with 2 LTCs and 39\% with 5 LTCs as the disk bandwidth becomes fully utilized.  
The same observation holds true with the W100 workload.

With Zipfian, the throughput does not scale.
85\% of requests reference keys assigned to the first LTC. 
The CPU of this LTC becomes fully utilized to dictate the overall system performance. 
Section~\ref{sec:eval:ltc-load-balancing} describes migration of a range from one LTC to another to balance load, enhancing performance.

\begin{figure}
\begin{subfigure}[t]{0.47\columnwidth}
\centering
\includegraphics[width=\textwidth]{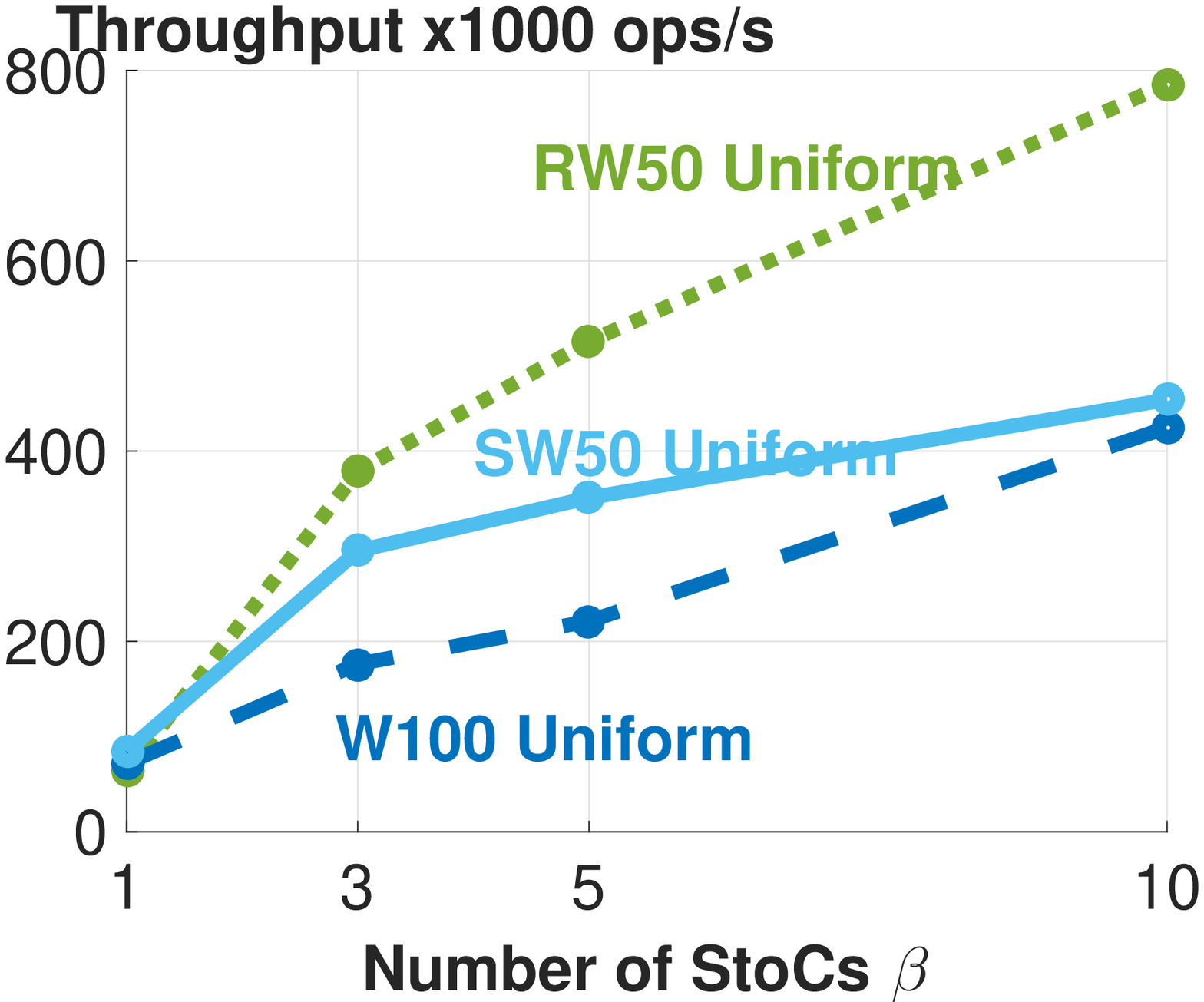}
\caption{Throughput.}
\label{fig:eval:5ltc-thpt-stocs-uniform}
\end{subfigure}
\quad
\begin{subfigure}[t]{0.47\columnwidth}
\centering
\includegraphics[width=\textwidth]{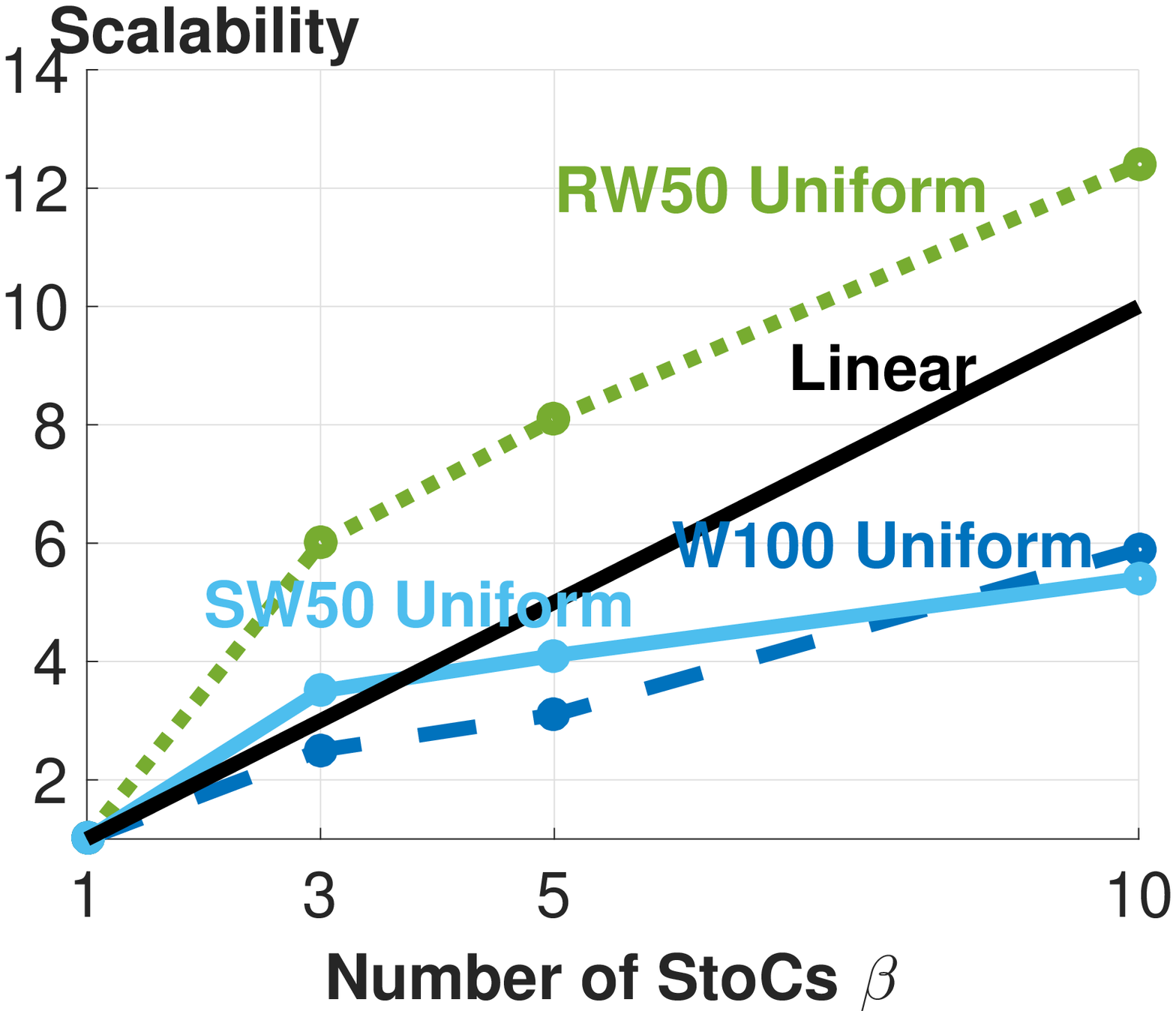}
\caption{Scalability.}
\label{fig:eval:5ltc-scale-stocs-uniform}
\end{subfigure}
\caption{Throughput and scalability as a function of $\beta$ with $\eta=5$, $\rho=1$, $\alpha=64$, and $\delta=256$.}
\label{fig:eval:5ltc-stocs-uniform}
\end{figure}

\noindent{\bf Scalability of 5 LTCs as a function of StoCs:}  
Figure~\ref{fig:eval:5ltc-stocs-uniform} shows the throughput and scalability of different workloads with 5 LTCs as we increase the number of StoCs from 1 to 10.  We show the results with Uniform.  With Zipfian, one LTC containing the popular keys becomes the bottleneck to dictate the overall processing capability.  With both RW50 and W100, this LTC bottleneck is encountered with 3 or more StoCs.  With SW50, the bottleneck LTC is present even with 1 StoCs.  (See Section~\ref{sec:eval:ltc-load-balancing} for re-organization of ranges across LTCs to address this bottleneck.)

The results with Uniform are very interesting.  We discuss each of RW50, SW50, and W100 in turn.  RW50 with Uniform utilizes the disk bandwidth of one StoC fully.
It benefits from additional StoCs, resulting in a higher throughput as we increase the number of StoCs.  
The size of the data stored on a StoC decreases from 20 GB with 1 StoC to 2 GB with 10 StoC.  
With a single StoC, the operating system's page cache is exhausted. 
With more StoCs, reads are served using the operating system's page cache, causing the throughput to scale super-linearly.

W100 scales up to 3 StoCs and then increases sublinearly with additional StoCs.  Writes stall since size of Level$_0$ is set to 2 GB for all StoC settings.  They constitute 91\%, 83\%, 79\%, 67\% of experiment time with 1, 3, 5, 10 StoCs, respectively.  This duration must decrease linearly in order for system throughput to scale linearly. 

SW50 with Uniform utilizes the CPU of all 5 LTCs fully with 3 StoCs.
Hence, increasing the number of StoCs provides no performance benefit, resulting in no horizontal scale-up.

\label{sec:eval:scalability}

\subsubsection{Load balancing across LTCs}\label{sec:eval:ltc-load-balancing}
\begin{table}
\small
\centering
\caption{Throughput (ops/s) with Zipfian and $\eta=5$, $\beta=10$, $\omega=64$, $\alpha=4$, $\delta=8$, $\rho=1$.}
\begin{tabular}{cccc} \hline
Workload & Before migration & After migration & Improvement  \\ \hline
RW50 & 415,798 & 988,420 & 2.38 \\ \hline
SW50 & 50,396 & 210,193 & 4.17 \\ \hline
W100 & 298,677 & 503,230 & 1.68 \\ \hline
\end{tabular}
\label{tab:eval:load-balancing}
\end{table}

The coordinator may monitor the utilization of LTCs and implement a load balancing technique such as~\cite{accordian-elastic,kairos}.  
With Nova-LSM, it is trivial to migrate ranges across LTCs (detailed in Section~\ref{sec:elasticity}). 
This section focuses on experimental results of Figure~\ref{fig:eval:5ltc-stocs-uniform} with a Zipfian distribution of access that caused a single LTC to become the bottleneck while the utilization of other LTCs is approximately 20\%.  
We migrate ranges across the LTCs to approximate a balanced load across different LTCs.  The migration requires only a few seconds because the bottleneck LTC pushes its ranges to four different LTCs and each LTC uses one RDMA READ to fetch log records from the StoC for each the relevant memtable.  

Once the migration completes, both the throughput and scalability improve dramatically, see Table~\ref{tab:eval:load-balancing}.  
With both W100 and SW50 using Zipfian, the scalability is similar to that shown with Uniform, see Figure~\ref{fig:eval:5ltc-scale-stocs-uniform}.  
Moreover, the explanation for why their throughput does not scale linearly is also the same as the discussion with Uniform.  

With RW50 using Zipfian, the scalability improves from 3x with 10 StoCs to 6x after migration.  
RW50 with Zipfian observes cache hits in memtables, utilizing the CPU of 5 LTCs fully.  Hence, similar to SW50, it does not utilize the disk bandwidth of all 10 StoCs.

\begin{figure}%
\begin{subfigure}[t]{0.47\columnwidth}
\includegraphics[width=\textwidth]{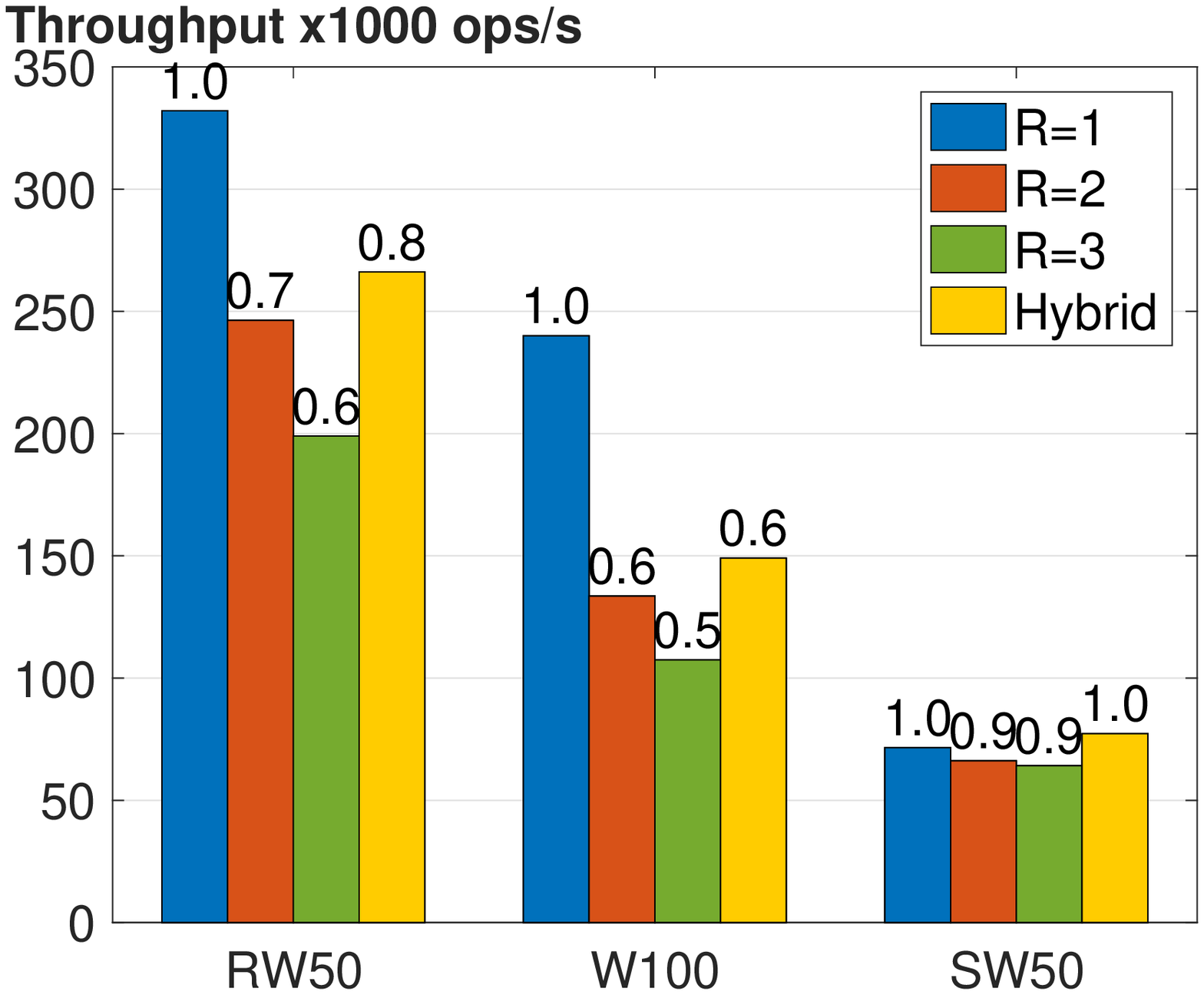}
    \caption{Throughput.}
    \label{fig:eval:nova-lsm-sstable-replication-thpt}
    \end{subfigure}%
\quad
\begin{subfigure}[t]{0.47\columnwidth}
\includegraphics[width=\textwidth]{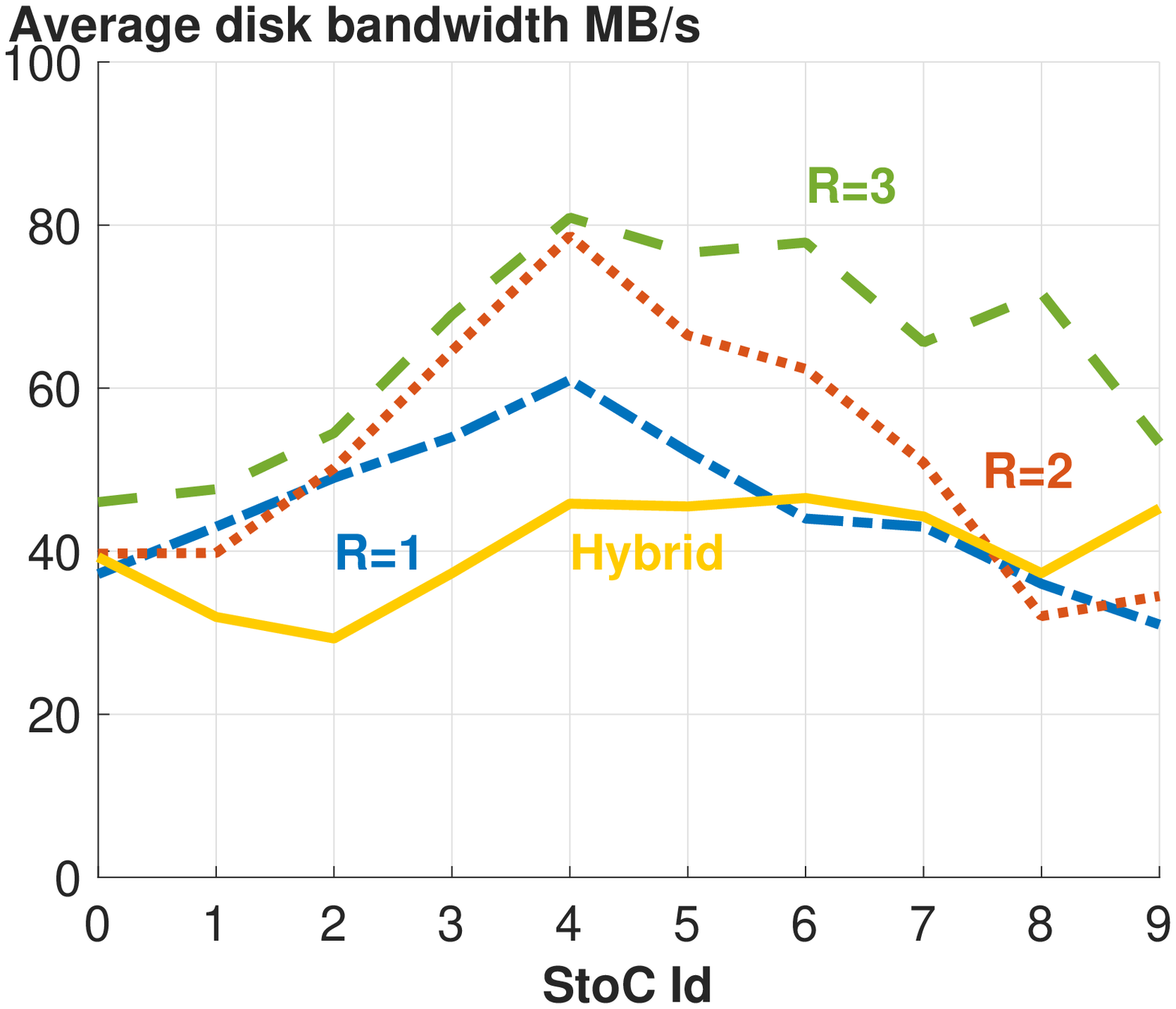}
    \caption{Disk bandwidth of W100.}
    \label{fig:eval:nova-lsm-sstable-replication-disk}
\end{subfigure}%
\caption{Impact of SSTable replication $R$ with Uniform and $\eta=1$, $\beta=10$, $\alpha=64$, $\delta=256$.}
\label{fig:eval:nova-lsm-sstable-replication}
\end{figure}

\subsubsection{SSTable Replication}\label{sec:eval:sstable-replication}
An LTC replicates an SSTable and its fragments to continue operation in the presence of failures that render a StoC unavailable. 
Figure~\ref{fig:eval:nova-lsm-sstable-replication} shows the throughput of different workloads as a function of the degree of replication  for a SSTable, $R$. 
With hybrid, a SSTable contains one parity block for its 3 data block fragments and 3 replicas of its metadata blocks\footnote{While a metadata block is approximately 200 KB in size, a data block fragment is approximately 5.3 MB in size.}.
Replicating a SSTable consumes disk bandwidth, reducing the overall system throughput. 
Its impact is negligible with a workload that utilizes the CPU fully, e.g., SW50 workload of Figure~\ref{fig:eval:nova-lsm-sstable-replication-thpt}.
With a disk intensive workload, say W100, the throughput almost halves when we increase $R$ from one to two.  
It drops further from two to three replicas.
However, it is not as dramatic because we use power-of-d.  
As shown in Figure~\ref{fig:eval:nova-lsm-sstable-replication-disk}, the overall disk bandwidth utilization with $R=2$ is uneven with StoC 4 showing a high utilization.  
With $R=3$, other StoCs start to become fully utilized (due to power-of-d), resulting in a more even utilization of StoCs.  Hence, the drop in throughput is not as dramatic as $R=2$. 

\begin{figure}%
\begin{subfigure}[t]{0.47\columnwidth}
\includegraphics[width=\textwidth]{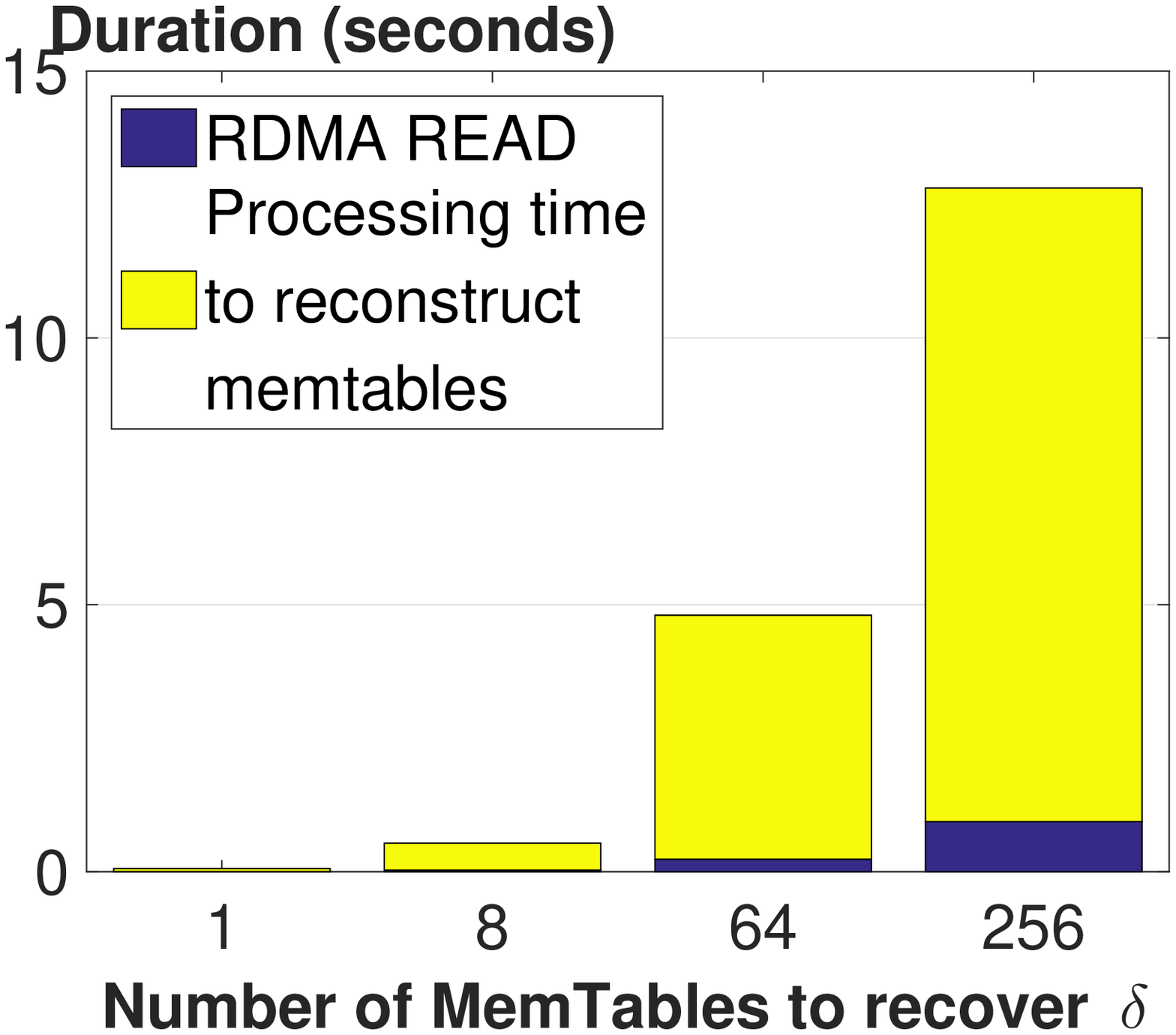}
    \caption{RDMA and CPU processing time with 1 recovery thread.}
    \label{fig:eval:recovery-duration}
    \end{subfigure}%
\quad
\begin{subfigure}[t]{0.47\columnwidth}
\includegraphics[width=\textwidth]{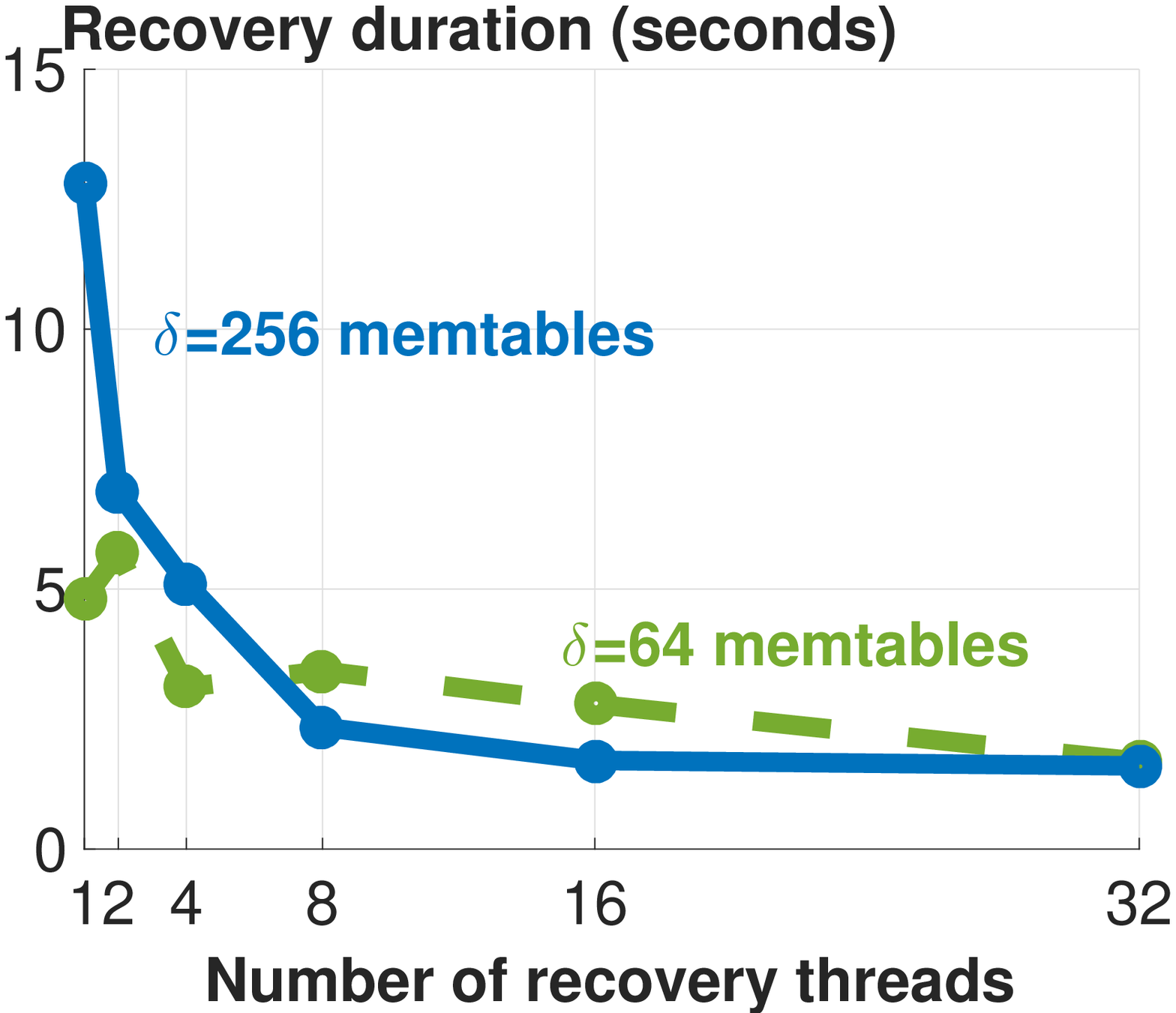}
    \caption{Recovery duration.}
    \label{fig:eval:recovery-thread}
\end{subfigure}%
\caption{Recovery.}
\label{fig:eval:recovery}
\end{figure}

\subsubsection{Recovery Duration}\label{sec:eval:recovery}
When an LTC fails, Nova-LSM recovers its memtables by replaying their log records from in-memory StoC files. 
The recovery duration is dictated by the memtable size, the number of memtables, and the number of recovery threads. 
We first measure the recovery duration as a function of the number of memtables. 
Figure~\ref{fig:eval:recovery-duration} shows the recovery duration increases as the number of memtables increases.
We consider different numbers of memtables on the x-axis because given a system with $\eta$ LTCs, the ranges of a failed LTC may be scattered across $\eta$-1 LTCs.  
This enables reconstruction of the different ranges in parallel, by requiring different LTCs to recover different memtables assigned to each range.

An LTC fetches all log records from one StoC at the line rate using RDMA READ. 
It fetches 4 GB of log records in less than 1 second.
We also observe that reconstructing memtables from log records dominates the recovery duration. 

A higher number of threads speeds up recovery time significantly, see Figure~\ref{fig:eval:recovery-thread}. 
With 256 memtables, the recovery duration decreases from 13 seconds to 1.5 seconds as we increase the number of recovery threads from 1 to 32. 
With 32 recovery threads, the recovery duration is dictated by the speed of RDMA READ. 

\subsection{Comparison with Existing Systems}\label{sec:eval:leveldb}
This section compares Nova-LSM with LevelDB and RocksDB given the same amount of hardware resources.  
We present results from both a single node and a ten-node configuration using different settings: 
\begin{itemize}[leftmargin=*]
\small
    \item \textbf{LevelDB:} One LevelDB instance per server, $\omega$=1, $\alpha$=1, $\delta$=2.
    \item \textbf{LevelDB*:} 64 LevelDB instances per server, $\omega$=64, $\alpha$=1, $\delta$=2.
    \item \textbf{RocksDB:} One RocksDB instance per server, $\omega$=1, $\alpha$=1, $\delta$=128. 
    \item \textbf{RocksDB*:} 64 RocksDB instances per server, $\omega$=64, $\alpha$=1, $\delta$=2.
    \item \textbf{RocksDB-tuned:} One RocksDB instance per server with tuned knobs, $\omega=1$. We enumerate RocksDB knob values and report the highest throughput. Example knobs include size ratio, Level$_1$ size, number of Level$_0$ files to trigger compaction, number of Level$_0$ files to stall writes.
\end{itemize}

\begin{figure}
\begin{subfigure}[t]{0.5\textwidth}
\centering
\includegraphics[width=\textwidth]{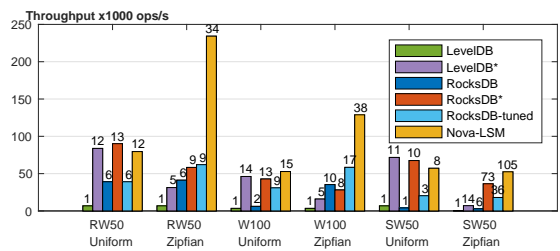}
\caption{10 GB database and 1 server.}
\label{fig:eval:nova-leveldb-10gb}
\end{subfigure}

\begin{subfigure}[t]{0.5\textwidth}
\centering
\includegraphics[width=\textwidth]{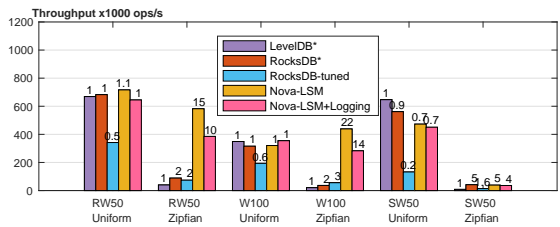}
\caption{100 GB database range partitioned across 10 servers.}
\label{fig:eval:nova-leveldb-100gb}
\end{subfigure}

\begin{subfigure}[t]{0.5\textwidth}
\centering
\includegraphics[width=\textwidth]{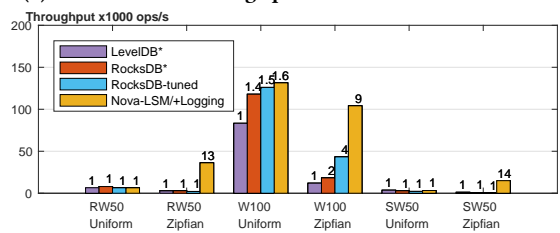}
\caption{1 TB database range partitioned across 10 servers.}
\label{fig:eval:nova-leveldb-1tb}
\end{subfigure}

\begin{subfigure}[t]{0.5\textwidth}
\centering
\includegraphics[width=\textwidth]{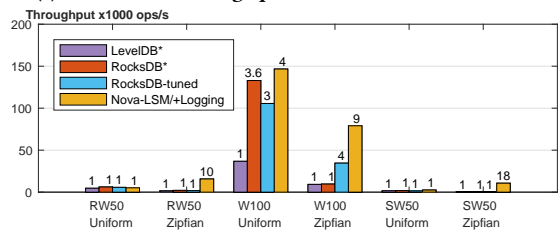}
\caption{2 TB database range partitioned across 10 servers.}
\label{fig:eval:nova-leveldb-2tb}
\end{subfigure}

\caption{Comparison of Nova-LSM with LevelDB and RocksDB.}
\label{fig:eval:nova-leveldb-10-servers}
\end{figure}

We conducted many experiments with different settings and observe similar results.  Here, we report results with logging disabled to provide an apple-to-apple comparison.  If we enable logging, performance of both LevelDB and RocksDB would decrease by a factor of 50.  For example, in Figure~\ref{fig:eval:nova-leveldb-10gb}, performance of LevelDB* would reduce from 50K to 1K with W100 Uniform.
We compare the different systems using 1 node and 10 nodes in turn.

Nova-LSM uses one range per server with 64 active memtables and 128 memtables per range, i.e., $\alpha=64$ and $\delta=128$.
Its logging is also disabled unless stated otherwise. 
With one server, a StoC client at LTC writes SSTables to its local disk directly.  
With 10 servers, each LTC scatters a SSTable across $\rho$=3 StoCs with different SSTables scattered across all 10 StoCs using power-of-6. 

Figure~\ref{fig:eval:nova-leveldb-10-servers} shows the comparison with three different database sizes using 1 node and 10 nodes.  We discuss each configuration in turn.

\subsubsection{One Node}
Figure~\ref{fig:eval:nova-leveldb-10gb} shows the throughput of LevelDB, RocksDB, and Nova-LSM with the aforementioned configurations.  
The x-axis of this figure identifies different workloads and access patterns.  
The numbers on top of each bar denote the factor of improvement relative to one LevelDB instance.  
Clearly, Nova-LSM achieves comparable performance with Uniform and outperforms both LevelDB and RocksDB with Zipfian.

\begin{table*}
\small
\centering
\caption{Response time (ms) with Zipfian.}
\begin{tabular}{cccc|ccc|ccc|ccc}
\hline
\multirow{2}{*}{}  & \multicolumn{3}{c}{R100} & \multicolumn{3}{c}{RW50} & \multicolumn{3}{c}{SW50} & \multicolumn{3}{c}{W100}\\\cline{2-13}
& Avg & p95 & p99 & Avg & p95 & p99 & Avg & p95 & p99 & Avg & p95 & p99 \\ \hline
LevelDB*&	 55 &	304 &	596 & 57 &	324 & 	586 & 118 &	615 & 1060 & 5.11 &	0.67 &	0.89 \\ \hline
RocksDB*& 56 &	272 & 742 &60 &	353 &	630  &128 &	650 &	984  &	3.97 &	0.69 &	1.08 \\ \hline
RocksDB-tuned &71 &	401	&667 & 75 &	381 & 	649  &139 &	769 &	1232 &	1.93 &	0.70 & 	1.02 \\ \hline
Nova-LSM  & 8 &	43 &	145 &9 &	61 & 	167 & 22 &	98 &	386 &	0.46 &	0.66 &	0.67 \\ \hline
\end{tabular}
\label{tab:eval:response-time}
\end{table*}

With Zipfian, Nova-LSM outperforms both LevelDB and RocksDB for all workloads.    
With RW50, Nova-LSM achieves 34x higher throughput than LevelDB with $\omega=1$ and 7x higher throughput when $\omega=64$. 
This is because use of Dranges enables Nova-LSM writes less data to disk, 65\% less. 
Nova-LSM merges memtables from Dranges with less than 100 unique keys into a new memtable without flushing them to disk. 
It uses the otherwise unutilized disk bandwidth to compact SSTables produced by different Dranges. 

With SW50 and Zipfian, the throughput of Nova-LSM is 105x higher than LevelDB with $\omega=1$ and 8x higher when $\omega=64$. 
While LevelDB/RocksDB scans through all versions of a hot key, Nova-LSM seeks to its latest version and skips all of its older versions. 
However, maintaining the range index incurs a 15\% overhead with Uniform since the CPU is 100\% utilized, causing Nova-LSM to provide a lower throughput than LevelDB with $\omega=64$.

\subsubsection{Ten Nodes with Large Databases}\label{sec:eval:leveldb-large-db}
Figure~\ref{fig:eval:nova-leveldb-10-servers} shows a comparison of Nova-LSM with existing systems using 10 nodes and two different database sizes:  100 GB and 1 TB.  
With 1 TB, the LSM-tree has 4 levels for each range.
LevelDB* and RocksDB* result in a total of 640  instances.  
We analyze two different configurations of Nova-LSM with and without logging.

With Zipfian, Nova-LSM outperforms both LevelDB and RocksDB by more than an order of magnitude, see Figure~\ref{fig:eval:nova-leveldb-100gb} and Figure~\ref{fig:eval:nova-leveldb-1tb}.
With a 100 GB database, Nova-LSM provides 22x higher throughput than LevelDB*. 
The throughput gain is 9x with the 1 TB database. 
This is because 85\% of requests are referencing the first server, causing it to become the bottleneck. 
With LevelDB* and RocksDB*, the disk bandwidth of the first server is fully utilized and dictates the overall throughput. 
With Nova-LSM, while the CPU of the server hosting the LTC assigned the range with the most popular keys becomes the bottleneck, this LTC uses the bandwidth of all 10 disks. 
This also explains why the throughput is lower when logging is enabled. 

With Uniform, the throughput of Nova-LSM is comparable to other systems for RW50 and SW50. 
This is because the operating system's page caches are exhausted and the cache misses must fetch blocks from disk. 
These random seeks limit the throughput of the system. 
This also explains why the overall throughput is much lower with the 1 TB database when compared with the 100 GB. 
With W100, Nova-LSM provides a higher throughput because the 10 disks are shared across 10 servers and a server scatters a SSTable across 3 disks with the shortest queues.

\subsubsection{Response Times with a 2 TB Database}
% \begin{figure}
%     \centering
%     \includegraphics[width=\linewidth]{figs/nova_leveldb_10servers_2tb_response_time_no_r100.eps}
%     \caption{Comparison of Nova-LSM with LevelDB and RocksDB.}
%     \label{fig:eval:response-time}
% \end{figure}

We analyzed the average, p95 and p99 response time of the different systems with R100, RW50, SW50, and W100 workloads, see Table~\ref{tab:eval:response-time}.  
These experiments quantify response time with a low system load:  60 YCSB threads issuing requests to a 10 node Nova-LSM configuration.  
With Uniform, all systems provide comparable response times.  
With Zipfian, Nova-LSM outperforms the other systems more than 3x.  
% Table~\ref{tab:eval:response-time} shows the observed response times with R100, RW50, SW50, and W100 workloads.  
A fraction of reads reference keys not found in the index and must search the other levels.  This fraction results in a higher observed response times with p95 and p99.  
Nova-LSM enhances the average, p95, and p99 response times by using its index structures and all 10 disks to process concurrent reads and writes. 
With RocksDB and LevelDB, clients direct 85\% of requests to one disk, degrading response times due to queuing delays.

With W100, a surprising result is the higher average response time when compared with p95 and p99 for both LevelDB and RocksDB.  Write-stalls cause a few requests to encounter long queuing delays on 1 disk.  While the p95 and p99 calculations drop these outliers, calculation of the average includes them to result in a higher response time.

\begin{figure}
    \centering
    \includegraphics[width=\linewidth]{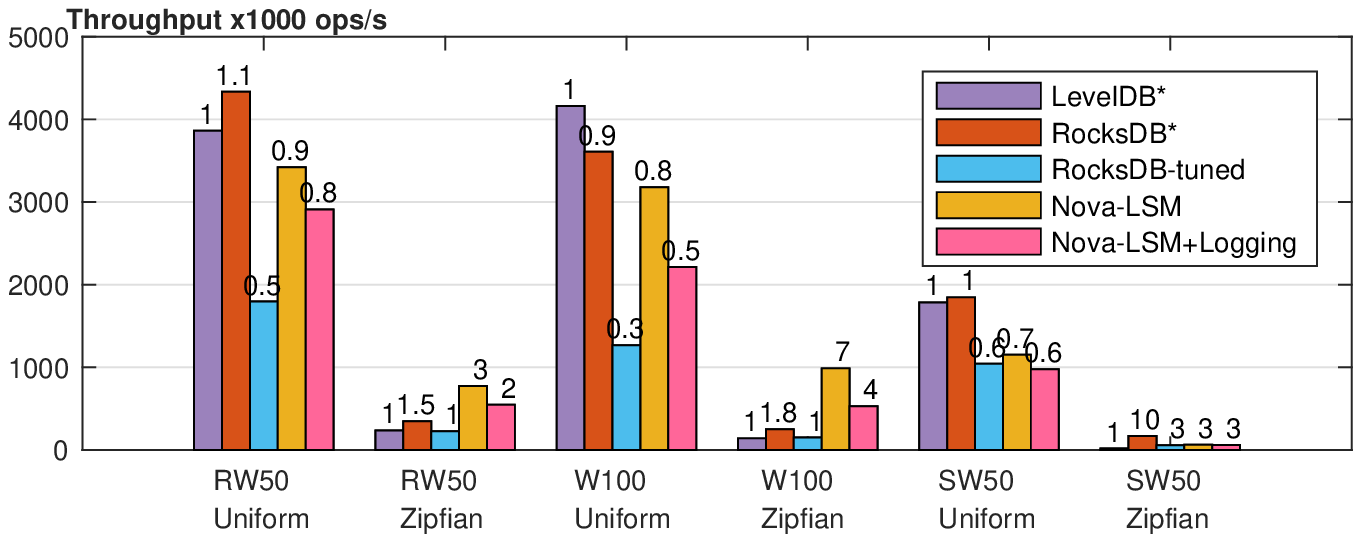}
    \caption{Comparison of Nova-LSM with LevelDB and RocksDB on tmpfs.}
    \label{fig:eval:tmpfs}
\end{figure}
\subsubsection{Faster Storage}
Reported trends hold true with storage devices faster than hard disk.  In its most extreme, the storage may be as fast as memory.  We emulated this by using an in-memory file system, tmpfs~\cite{tmpfs}, with a 100 GB database partitioned across 10 servers. In these experiments, the CPU is the bottleneck resource.  Nova-LSM continues to provide a higher throughput than LevelDB* with Zipfian.  This is a factor of 2 to 7 depending on whether Nova-LSM is configured with logging.  While Nova-LSM offloads compactions to StoCs, LevelDB and RocksDB use a single node to process most of requests and compactions.  With SW50 workload, RocksDB* provides a higher throughput than Nova-LSM by processing fewer memtables.  RocksDB* contains 64 application ranges per server with 2 memtables per range. A scan searches 2 memtables. With Nova-LSM, a scan may overlap multiple Dranges with 2 memtables assigned to each Drange.  In the reported experiments, a scan may search a maximum of 26 memtables.  With Uniform, Nova-LSM provides 10\% to 30\% lower throughput due to the maintenance of the index structures with writes and RDMA threads pulling for requests. Both are CPU intensive tasks.

\section{Elasticity}\label{sec:elasticity}

Section~\ref{sec:eval} highlights different components of Nova-LSM must scale elastically for different workloads.  While LTCs must scale elastically with CPU intensive workloads such as SW50, StoCs must scale elastically with disk I/O intensive workloads such as RW50.  

Elastic scalability has been studied extensively.  There exist heuristic-based techniques~\cite{scads2011,cake12,accordian-elastic,vertical16} and a mixed integer linear programming~\cite{accordian-elastic, kairos} formulation of the problem.  
The coordinator of Figure~\ref{fig:arch} may adopt one or more of these techniques.  We defer this to future work.  Instead, this section describes how the coordinator may add or remove either an LTC or a StoC from a configuration.  Scaling StoCs migrates data while scaling LTCs migrates metadata and in-memory memtables.  Both use RDMA to perform this migration.

\noindent{\bf Adding and Removing LTCs:} Scaling LTCs migrates one or more ranges from a source LTC to one or more destination LTCs.  It requires the source LTC to inform the destination LTC of the metadata of the migrating range.  This includes the metadata of LSM-tree, Dranges, Tranges, lookup index, range index, and locations of log record replicas.  It sends the metadata to the destination LTC using RDMA WRITE.  The destination LTC uses this metadata to reconstruct the range.  It also uses multiple background threads to reconstruct the memtables by replaying their log records in parallel.  In experiments reported below, the total amount of data transferred when migrating a range was 45 MB.  Approximately 1\% of this data, 613 KB, was for metadata.  The remaining 99\% were log records to reconstruct partially full memtables. 

\begin{figure}
\begin{subfigure}[t]{0.5\textwidth}
\centering
\includegraphics[width=\textwidth]{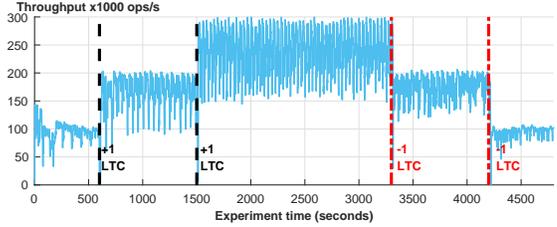}
\caption{SW50 Uniform, starting config has $\eta=1$, $\beta=13$, $\alpha=4$, $\delta=8$, $\omega=64$.}
\label{fig:eval:nova-elastic-sw50}
\end{subfigure}

\begin{subfigure}[t]{0.5\textwidth}
\centering
\includegraphics[width=\textwidth]{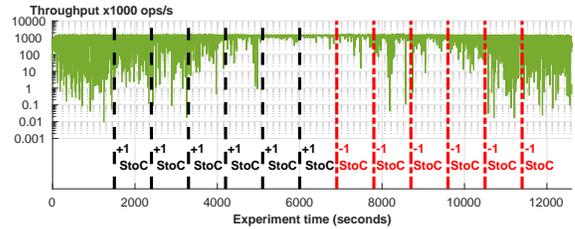}
\caption{RW50 Uniform, starting config has $\eta=3$, $\beta=3$, $\alpha=4$, $\delta=8$, $\omega=64$.}
\label{fig:eval:nova-elastic-rw50}
\end{subfigure}
\caption{Elasticity.}
\label{fig:eval:nova-elastic}
\end{figure}

The coordinator may redirect clients that reference a migrating range to contact the destination LTC as soon as it renders the migration decision.  The destination LTC blocks a request that references the migrating range until it completes reconstructing its metadata and relevant memtables.  (It may give a higher priority to reconstruct those memtables with pending requests.)  In our experiments, the maximum delay incurred to reconstruct metadata of a migrating range was 570 milliseconds.

Figure~\ref{fig:eval:nova-elastic-sw50} shows the throughput of SW50 workload with a base configuration consisting of 1 LTC and 13 StoCs.  This workload utilizes the CPU of the LTC fully, providing a throughput of 100K operations/second.  There is sufficient disk bandwidth to render write stalls insignificant, resulting in an insignificant throughput variation from 90K to 100K operations per second.  We add 1 new LTC and migrate half the ranges to it.  This almost doubles the throughput.  However, write stalls result in a significantly higher throughput variation ranging from 100K to 200K operations per second.  Finally, we add a 3rd LTC.  The peak throughput increases to 300K operations per second, utilizing the CPU of the 3rd LTC fully.  Write stalls cause the observed throughput to vary from 175K to 300K operations per second.  While the average throughput does not scale linearly with additional LTCs due to write stalls, the peak throughput increases linearly as a function of LTCs.

\noindent{\bf Adding and Removing StoCs:}
% \subsection{Adding and Removing StoCs}
When a StoC is added to an existing configuration, LTCs assign new SSTables to the new StoC immediately using power-of-d.  Nova-LSM also considers the possibility of this StoC having replicas of data as this StoC may have been a part of a previous configuration (and shut down due to a low system load). 
A file is {\em useful} if it is still referenced by a SSTable. Otherwise, it is obsolete and is deleted. 

Each file name maintains its range id and SSTable file number.  A new StoC starts by enumerating its files to retrieve each file's range id and SSTable file number. 
For each file, it identifies the LTC hosting its range and queries it to determine if the SSTable is still referenced.  If it is not referenced then the StoC deletes this file.  Otherwise, the LTC extends the SSTable metadata to reference this replica for future use.  This may increase the number of replicas for a data fragment, a metadata block, and a parity block beyond that specified by an application.  
If space is at a premium, this data may be deleted.  
Given the inexpensive price of disks (2 to 3 pennies per Gigabyte) and the immutable nature of SSTables, a StoC may maintain these additional replicas.  
They are useful because they minimize the amount of data that must be migrated once this StoC is shutdown during off-peak hours.  Once an LTC repairs metadata of a SSTable, it issues reads to the new StoC using power-of-d immediately.

When a StoC is shut down gracefully, each LTC analyzes its StoC's files.  It identifies those SSTables fragments or parity blocks pertaining to SSTables of its assigned ranges.  Next, it identifies a destination StoC for each fragment while respecting placement constraints, e.g., replicas of a SSTable must be stored on different StoCs.  Lastly, it informs the source StoCs to copy the identified files to their destinations using RDMA. 

% When a StoC is shutdown and added repeatedly, it is possible for a fragment to contain more replicas than its pre-specified threshold.  An LTC identifies these fragments and does not migrate them when shutting down a StoC.  

Figure~\ref{fig:eval:nova-elastic-rw50} shows the throughput of the RW50 workload with 3 LTCs and 3 StoCs.  Each SSTable is assigned to 1 StoC, $\rho$=1.  We increase the number of StoCs by one every 15 minutes until the number of StoCs reaches 9. With 8 StoCs, the number of write stalls is diminished significantly as there is sufficient disk bandwidth to keep up with compactions. The number of write stalls is further reduced with the 9th StoC.  Next, we remove the StoCs by 1 every 15 minutes until we are down to 3 StoCs.  This reduces the observed throughput as RW50 is disk I/O intensive.

In Figure~\ref{fig:eval:nova-elastic-rw50}, the average throughput increases from 700K with 3 StoCs to 1.35M with 9 StoCs.  This 2 fold increase in throughput is realized by increasing the number of StoCs 3 folds.  The sub-linear increase in throughput is because the load is unevenly distributed across 9 StoCs even though we use power-of-d.  
The average disk bandwidth utilization varies from 76\% to 93\%.
The base configuration does not have this limitation since it consists of 3 StoCs and each SSTable is partitioned across all 3 StoCs, $\rho=3$.

\section{Related Work}\label{sec:related-work}
\noindent\textbf{LSM-tree data stores:} 
Today's monolithic LSM-tree systems~\cite{rocksdb,hyperleveldb,cLSM,bLSM,bourbon,lsm-memory-wall} require a few memtables to saturate the bandwidth of one disk.
Nova-LSM uses a large number of memtables to saturate the disk bandwidth of multiple StoCs. 
Its Dranges, lookup index, and range index complement this design decision. 

TriAD~\cite{triad} and X-Engine~\cite{xengine} separate hot keys and cold keys to improve performance with data skew.
Nova-LSM constructs Dranges to mitigate the impact of skew. 
Dranges that contain hot keys maintain these keys in their memtables without flushing them to StoCs. 
Dranges also facilitate compaction of SSTables at Level$_0$. 

FloDB~\cite{flodb} and Accordion~\cite{accordion} redesign LSM-tree for large memory. 
Both do not address the challenges of compaction and expensive reads due to a large number of Level$_0$ SSTables. 
Nova-LSM maintains lookup index and range index to expedite gets and scans. 
With FloDB, a scan may restart many times and block writes. 
With Nova-LSM, a scan never restarts and does not block writes. 

An adaptive memory management technique for reads and writes referencing multiple LSM-Trees is described in~\cite{lsm-memory-wall}.  For writes to each LSM-tree, it maintains one active memtable and a set of immutable memtables organized into levels.  In contrast, Nova-LSM's LTC maintains multiple active memtables for a LSM-tree.  One for each Drange.  This enables different threads (CPU cores) to write to different Dranges concurrently.  An LTC does not flush immutable memtables with fewer than a pre-specified number of unique values (100 in our experiments) to disk.  With a Drange that consists of 3 or more immutable memtables, one may use write technique of~\cite{lsm-memory-wall}.  Quantifying the tradeoffs associated with these alternatives is a short term research direction.  Note that an LTC may use the memory tuner of~\cite{lsm-memory-wall} to manage its buffer cache and write memory. 

AC-Key is a caching technique to enhance performance of reads and scans for LSM-based key-value stores~\cite{ackey}.  Nova-LSM may implement AC-Key or CAMP~\cite{camp2014} as a component.

Both RocksDB subcompactions~\cite{rocksdb-subcompaction} and Vinyl~\cite{vinyl} use detective techniques while Nova-LSM's proposed Dranges is a preventive technique.  RocksDB may construct ranges at compaction time and assign a range to each sub compaction thread.  This is redundant with Nova-LSM because it uses Dranges. Vinyl splits on-disk SSTables into slices.  Nova-LSM splits memtables (of a range) using Dranges.  A Level$_0$ SSTable of Vinyl may span the entire keyspace of a range.  Dranges prevent this in Nova-LSM.

The compaction policy is intrinsic to the performance of an LSM-tree. 
Chen et al.~\cite{lsmtree-survey} present an extensive survey of alternative LSM-tree designs. 
% It also conducted an extensive evaluation of alternative policies. 
PebblesDB~\cite{pebblesdb} uses tiering to reduce write amplification at the cost of more expensive reads.
Dostoevsky~\cite{Dostoevsky} analyzes the space-time trade-offs of alternative policies and proposes new policies that achieve a better space-time trade-off.
Nova-LSM may use these policies as its future extensions.

Several studies redesign LSM-tree to use fast storage mediums, SSD and NVM~\cite{wisckey, novelsm,slm-db,matrixkv2020}. 
NoveLSM reduces write amplification by placing memtables on NVM and updating the entries in-place~\cite{novelsm}.
MatrixKV uses a multi-tier DRAM-NVM-SSD to reduce write stalls and mitigate write amplifications~\cite{matrixkv2020}.  Nova-LSM's StoC may use these storage mediums and techniques.

Multi-node data stores use LSM-tree to store their data, e.g., BigTable~\cite{bigtable}, Cassandra~\cite{cassandra,lsm-cassandra-compaction}, and AsterixDB~\cite{asterixdb}.
Nova-LSM is different because it separates its compute from storage.
Its LTC also offloads compaction to StoCs.

% \noindent\textbf{Edge-cloud LSM-tree data stores:} 
Cooperative LSM~\cite{coolsm}, CooLSM, is designed for wide-area networks where processing and storage span both edge and cloud nodes.  Its architecture addresses the network latency associated with geographically distributed applications such as IoT.  It consists of Ingestor processes that receive the write requests and maintain Levels L0 and L1, range partitioned Compactor processes that maintain the rest of levels, and Reader processes that maintain a copy of the entire LSM tree for recovery and read availability.   Nova-LSM and its components are different as they are designed for a data center with fast network connectivity.  Hence, data is range partitioned across LTCs that maintain only memtables.  LTCs may off-load compaction to StoCs making them similar to CooLSM’s compactors.  However, Nova-LSM does not range partition data across StoCs.  

\noindent\textbf{Separation of storage from processing:} 
A recent trend in database community is to separate storage from processing. 
Example databases are Aurora~\cite{aurora}, Socrates~\cite{socrates}, SolarDB~\cite{solardb}, HBase~\cite{hbase}, Taurus~\cite{Taurus}, and Tell~\cite{tell}.
Nova-LSM is inspired by these studies.  
Our proposed separation of LTC from StoC, declustering of a SSTable to share disk bandwidth, parity-based and replication techniques for data availability, and power-of-d using RDMA are novel and may be used by the storage component of these prior studies.

An LSM-tree may run on a distributed file system to utilize the bandwidth of multiple disks. 
Orion~\cite{orion} is a distributed file system for non-volatile memory and RDMA-capable networks. 
Hailstorm~\cite{Hailstorm} is designed specifically for LSM-tree data stores. 
POLARFS~\cite{polardb} provides ultra-low latency by utilizing networking stack and I/O stack in userspace.
Nova-LSM also harnesses the bandwidth of multiple disks. 
Its LTC scatters a SSTable across multiple StoCs and uses power-of-d to minimize delays.%%%concurrent writes on a StoC. 

Rockset~\cite{rockset} separates processing from storage by offloading compaction to Amazon S3.  
Nova-LSM uses off-the-shelf hardware and software, controlling the number of physical storage devices ($\rho$) that store a SSTable.

\noindent\textbf{RDMA-based systems:} 
LegoOS~\cite{legoos} disaggregates the physical resources of a monolithic server into network-attached components. 
These components exchange data using RDMA.
Nova-LSM may implement its components using LegoOS.  For example, StoC may be implemented using LegoOS's file system.

Tailwind~\cite{tailwind} and Active-Memory~\cite{active-memory} are replication techniques designed for RDMA. 
Both replicate data using RDMA WRITE.
LogC also uses RDMA WRITE to replicate log records for high availability. 

Several studies present optimizations on the use of RDMA to enhance performance and scalability for in-memory key-value stores~\cite{rdma-occ,farm,rdma-eval,skeda}.
Nova-LSM is a disk-based data store. 
Its components use dedicated threads to minimize the number of QPs.

\noindent
{\bf In-memory key-value stores:}
In-memory KVSs (FaRM and its extension with transactions~\cite{farm,farm15}) are a different class of systems than persistent KVSs (Nova-LSM).  Their target different applications.  In-memory KVSs outperform persistent KVSs because DRAM is significantly faster than SSD/HDD. 
FASTER~\cite{faster} uses an in-memory hash table to process gets, puts, and RMW expeditiously.  It states that its targets applications are different than KVSs based on LSM-tree and does not support a sorted order of keys for processing scans.

\section{Future Work}\label{sec:conclusion}
% We are extending Nova-LSM in several exciting directions.  
% % First, we are developing an on-line data re-organization technique that migrates ranges from highly loaded LTC to one with a lower load to balance the load more evenly across the LTCs.  
% First, Section~\ref{sec:eval:horizontal-scale} shows  
% Nova-LSM is highly elastic.  We plan to extend the coordinator of Figure~\ref{fig:arch} with techniques of~\cite{accordian-elastic, kairos,scads2011,cake12,vertical16,cloudscale2011} to decide the number of LTCs and StoCs, and what ranges and data fragments should be migrated.
% This coordinator may use the concept of Dranges to dynamically range partition data across LTCs, freeing the application from specifying ranges.
% %%%We are exploring the feasibility of an online planner that monitors a deployment and adjusts its number of LTCs and StoCs dynamically to realize the highest price/performance for a workload.
% Second, we plan to revisit the concept of Dranges and the alternative ways of reorganizing them to balance their load (Section~\ref{sec:dranges}).  There are several possibilities and this study focused on one design only.  Understanding the tradeoffs associated with the alternatives further enhances the reported performance benefits of Nova-LSM.
% Finally, we plan to build components that use interfaces of LTC, LogC, and StoC to implement alternative data models, query, and transaction processing capabilities~\cite{nova}.

We are extending Nova-LSM with new functionalities, application Service Level Agreements (SLAs), and cost analysis.  
An example functionality is a write batch that performs several puts and deletes atomically.   
We are exploring both lock-free and lock-based techniques to support this functionality.
%%%%We are investigating both lock-free and lock-based techniques to achieve strong consistency for multi-gets and scans across multiple LTCs.  
With the former, each key impacted by a write batch is extended to identify the other keys and sequence numbers in the batch.  
A read that fetches the value of two or more of these keys detects the metadata and verifies their sequence numbers to ensure it does not observe values of a concurrent write batch. 
Otherwise, it aborts and restarts.
A lock-based technique in combination with a two-phase commit protocol may implement strict serial schedules with write batch, multi-get, and scan that span multiple LTCs.

SLAs include performance, availability, and consistency requirements of an application.  
SLAs motivate an auto-tuner that scales components both vertically~\cite{vertical16} and horizontally~\cite{accordian-elastic}.  
This must be done intelligently per discussions of Figure~\ref{fig:write-stall-intro}.

Finally, a cost-analysis quantifies the expense associated with the hardware platform, power usage, and maintenance.   While networking hardware such as RDMA is expensive, should it reduce the size of a system 10 folds then its savings in power and space may outweigh its expenses.

%ACKNOWLEDGMENTS are optional
\section{Acknowledgments}
We gratefully acknowledge use of CloudLab network testbed~\cite{cloudlab} for all experimental results presented in this paper.
We thank SIGMOD's anonymous reviewers for their valuable comments. 

\bibliographystyle{ACM-Reference-Format}
\bibliography{main}  % vldb_sample.bib is the name of the Bibliography in this case

\end{document}